\documentclass[fleqn,usenatbib]{mnras}

\usepackage{newtxtext,newtxmath}

\usepackage[T1]{fontenc}
\usepackage{subcaption}
\DeclareRobustCommand{\VAN}[3]{#2}
\let\VANthebibliography\thebibliography
\def\thebibliography{\DeclareRobustCommand{\VAN}[3]{##3}\VANthebibliography}

\usepackage{siunitx}
\usepackage{soul,xcolor}
\usepackage{color,soul,natbib}
\usepackage{graphicx}	
\usepackage{amsmath}	
\usepackage{longtable}
\usepackage{mathtools}

\usepackage{ulem}
\usepackage{lineno}
\usepackage{indentfirst}

\title[Tidal Evolution Simulation]{The Application of Machine Learning in Tidal Evolution Simulation of Star-Planet Systems}

\author[Shuaishuai Guo]{
	Shuaishuai Guo$^{1,2,3,5}$,
	Jianheng Guo$^{1,2,3,5}$\thanks{E-mail: guojh@ynao.ac.cn},
	KaiFan Ji$^{1,4}$,
	Hui Liu$^{1,4}$,
	and Lei Xing$^{1,2,3,5}$
	\\
	$^{1}$Yunnan Observatories, Chinese Academy of Sciences, P.O. Box 110, Kunming 650011, People's Republic of China\\
	$^{2}$School of Astronomy and Space Science, University of Chinese Academy of Sciences, Beijing, People's Republic of China\\
	$^{3}$Key Laboratory for the Structure and Evolution of Celestial Objects, CAS, Kunming 650011, People's Republic of China\\
	$^{4}$Yunnan Key Laboratory of SolarPhysics and Space Science,650216,China\\
	$^{5}$International Centre of Supernovae, Yunnan Key Laboratory, Kunming 650216, P. R. China
}

\date{Accepted XXX. Received YYY; in original form ZZZ}

\pubyear{2024}

\begin{document}
		
		\label{firstpage}
		\pagerange{\pageref{firstpage}--\pageref{lastpage}}
		\maketitle
		
		\begin{abstract}
			With the release of a large amount of astronomical data, an increasing number of close-in hot Jupiters have been discovered. Calculating their evolutionary curves using star-planet interaction models presents a challenge. To expedite the generation of evolutionary curves for these close-in hot Jupiter systems, we utilized tidal interaction models established on MESA to create 15,745 samples of star-planet systems and 7,500 samples of stars. Additionally, we employed a neural network (Multi-Layer Perceptron - MLP) to predict the evolutionary curves of the systems, including stellar effective temperature, radius, stellar rotation period, and planetary orbital period. The median relative errors of the predicted evolutionary curves were found to be 0.15$\%$, 0.43$\%$, 2.61$\%$, and 0.57$\%$, respectively. Furthermore, the speed at which we generate evolutionary curves exceeds that of model-generated curves by more than four orders of magnitude.
			We also extracted features of planetary migration states and utilized lightGBM to classify the samples into 6 categories for prediction. We found that by combining three types that undergo long-term double synchronization into one label, the classifier effectively recognized these features. Apart from systems experiencing long-term double synchronization, the median relative errors of the predicted evolutionary curves were all below 4$\%$. Our work provides an efficient method to save significant computational resources and time with minimal loss in accuracy. This research also lays the foundation for analyzing the evolutionary characteristics of systems under different migration states, aiding in the understanding of the underlying physical mechanisms of such systems. Finally, to a large extent, our approach could replace the calculations of theoretical models.
		\end{abstract}
		
		
		
		\onecolumn

\section{Introduction} \label{sec:intro}
Since the discovery of the first hot Jupiter, 51 Pegasi b \citep{1995Natur.378..355M}, research on close-in hot Jupiters has never ceased. Among them, tidal interactions play a crucial role in the evolution of close-in hot Jupiters and their solar-like host stars \citep{2013LNP...861..301Z,2014ARA&A..52..171O,2019EAS....82....5M}. Orbital migration of the planet is accompanied by angular momentum transfer between the star and the planet, resulting in changes in the stellar rotation period and the semi-major axis of the planet's orbit \citep{2008ApJ...678.1396J,2009ApJ...692L...9L,2010ApJ...712.1107G,2016CeMDA.126..275B,2020MNRAS.498.2270B,2021MNRAS.508.3408L,2023MNRAS.520.3749L}. The stellar magnetic braking also leads to a gradual loss of stellar angular momentum \citep{2017ewas.confE...6S,2019A&A...621A.124B,2021A&A...651A...3A}.

Stellar rotation is a fundamental physical quantity, and for solar-like stars with long main-sequence lifetimes, their luminosity and effective temperature evolve slowly during the main sequence. The age determination of stars usually has an uncertainty of over 50$\%$ based on isochronal ages alone, making accurate stellar age estimation a challenging task. However, stellar age plays a crucial role in stellar structure evolution as well as the formation and evolution of exoplanets. Fortunately, there exists a correlation between stellar age and surface rotation for solar-like stars. \citet{1972ApJ...171..565S} first described a simple relationship between stellar rotation rate and age: $\Omega \propto t^{-0.5}$. Building upon Skumanich's work, \citet{2003ApJ...586..464B,2007ApJ...669.1167B} proposed a method called gyrochronology to estimate stellar ages by measuring the rotation period and color of low-mass stars. However, this method assumes that stellar rotation is not influenced by external factors. In the case of close-in hot Jupiter systems, the angular momentum transferred by orbital migration can clearly affect the stellar rotation period. \citet{2011A&A...525A..68B} first discovered that stars with planets rotate faster than stars without planets. \citet{2014MNRAS.442.1844B} studied a sample of 68 hot Jupiter host stars and found that gyrochronology yields younger ages compared to isochronal ages. Additionally, the migration state of hot Jupiters can be broadly categorized into three types: inward tidal migration ($\Omega < n$), outward tidal migration ($\Omega > n$), and double synchronization ($\Omega = n$). The first two cases are relatively straightforward, but for double synchronization, \citet{1980A&A....92..167H} used Lagrange multipliers to rigorously prove that, under the constraint of total angular momentum conservation, there is only one possible equilibrium state with characteristics of coplanarity, orbital circularization, and co-rotation (i.e., double synchronization). \citet{2015A&A...574A..39D} further considered the case of stellar angular momentum loss and proposed the existence of a pseudostable form in such systems.

In recent years, the tidal interaction between stars and planets has emerged as a popular research direction, resulting in complex evolutionary patterns within star-planet systems. On one hand, variations in stellar mass and metallicity lead to significant differences in stellar magnetic braking and evolution, which indirectly influence the stellar rotation speed and planetary orbital migration. On the other hand, the initial parameters of the planets can conversely affect orbital migration and changes in the stellar rotation rate. Currently, theoretical studies of star-planet tidal interactions are primarily based on physical modeling \citep{2013MNRAS.429..613O,2015A&A...580L...3M,2021MNRAS.508.3408L,2023MNRAS.520.3749L} followed by numerical simulations \citep{2023RAA....23i5014G,2024MNRAS.529.2893G}. In the numerical simulation models mentioned above, \citet{2013MNRAS.429..613O} conducted hydrodynamical simulations in convective spherical shells, while \citet{2015A&A...580L...3M} utilized the stellar evolution code (Starevol). Similar to \citet{2021MNRAS.508.3408L,2023MNRAS.520.3749L}, our preliminary work involved establishing interaction models based on the Modules for Experiments in Stellar Astrophysics (MESA). In \citet{2023RAA....23i5014G} and \citet{2024MNRAS.529.2893G}, we conducted larger-scale simulations and extensively explored the physical processes of double synchronous samples. We identified long-term double synchronous samples within the parameter space, providing training data for the classification task in this study. However, generating models through numerical simulations demands extensive computational resources and time. For example, using MESA to generate a single model requires an average of at least 20 minutes (CPU with 2.90 GHz main frequency and four cores). Fortunately, the application of machine learning has propelled astronomical research forward significantly. We are the first to apply machine learning methods to the tidal interactions between the close-in hot Jupiters and solar-like stars. This paper intends to combine machine learning with theoretical model-derived evolutionary curves to predict the evolutionary trajectories of star-planet systems and classify the migration states of planets.

Our work is divided into two main parts. Firstly, we employ the machine learning method MLP to perform regression on the stellar effective temperature ($logT_\mathrm{eff}$), stellar radius ($R_*$), stellar rotation period ($P_{\mathrm{rot}}$) and planetary orbital period ($P_{\mathrm{orb}}$). They are obtained from the tidal interaction model outputs of \citet{2023RAA....23i5014G} established within MESA (
In the text, we uniformly refer to the evolutionary curves generated by these four parameters as "evolutionary curves."). Secondly, we classify our samples into six types based on the migration states of the planets, and use lightGBM to predict and evaluate the classification performance. The structure of our paper is as follows: In Section \ref{sec:Phy}, we described the tidal interaction model and the pseudo-stable model we established in MESA. Section \ref{sec:data} outlined the generation and labeling methods for our regression and classification datasets. In Section \ref{sec:MAM}, we introduced the neural network MLP for regression problems and various classification methods for the classification task. Moving on to Section \ref{sec:result}, we presented the overall errors of the machine learning models for both regression and classification problems, and we also discussed the errors of individual systems regarding initial parameters and planetary migration states, as well as the speed of generating evolutionary curves. In the last section, we presented our conlusion and discussions.

\section{Physical Model} \label{sec:Phy}
Our model is based on Modules for Experiments in Stellar Astrophysics (MESA), version 11554 \citep{ 2011ApJS..192....3P,2013ApJS..208....4P,2015ApJS..220...15P,2018ApJS..234...34P,2019ApJS..243...10P}. In this work, we utilized MESA r11554 to construct our models. The implementation of stellar rotation in MESA can be found in \citet{2013ApJS..208....4P}, with updates provided in \citet{2019ApJS..243...10P}. Additionally, for solar-like stars under study, magnetic braking, which is significant, lacks a built-in interaction model in MESA's MESA\_star. Therefore, to incorporate magnetic braking, orbital evolution, and angular momentum transfer models into MESA, we employed the "run\_star\_extras.f" program file. We opted for the interface called "other\_torque", which is invoked within the code of "solve\_omega\_mix". This interface allows for the addition of extra torque to the stellar rotation model. The specific details of the model are as follows:

\subsection{Star-planet interaction model} \label{sec:style}
The model we used is based on the work of \citet{2023RAA....23i5014G}, assuming that planets are on circular orbits and tidally locked to their host stars, treated as point masses. We implemented the magnetic braking model into the rotational scheme provided by MESA, and incorporated models for planetary orbit evolution and angular momentum transfer. The magnetic braking model, planetary orbit evolution model, and stellar angular velocity evolution model we employed are as follows:

We use magnetic braking model of \citet{2015ApJ...799L..23M} scheme as following:

\begin{equation}
	\frac{\mathrm{d}L_{wind}}{\mathrm{d}t}=-T_0(\frac{\tau_{\mathrm{cz}}}{\tau_{\mathrm{cz\odot}}})^p(\frac{\Omega}{\Omega_\odot})^{p+1}		(unsaturated),
	\label{eq1}
\end{equation}

\begin{equation}
	\frac{\mathrm{d}L_{\mathrm{wind}}}{\mathrm{d}t}=-T_0{\chi}^p(\frac{\Omega}{\Omega_\odot})		(saturated),
	\label{eq2}
\end{equation}

\begin{equation}
	T_0=K(\frac{R_{*}}{R_\odot})^{3.1}(\frac{M_{*}}{M_\odot})^{0.5}{\gamma}^{2m},
	\label{eq3}
\end{equation}

\begin{equation}
	\gamma=\sqrt{1+(u/0.072)^2},
	\label{eq4}
\end{equation}
\begin{equation}
	u=\frac{\Omega}{\Omega_{\mathrm{crit}}},\Omega_{\mathrm{crit}}=\sqrt{\frac{GM_*}{{R_*}^3}},
	\label{eq5}
\end{equation}
where $L_{\mathrm{wind}}$ denotes the angular momentum lost by the magnetic stellar wind, while $R_*$, $\Omega$, and $\Omega_{\mathrm{crit}}$ represent the stellar radius, stellar rotation angular velocity, and critical angular rotation rate, respectively. The parameters $K$, $m$, $p$, and $\chi$ are constants. The equations (\ref{eq1}) and (\ref{eq2}) exhibit saturated and unsaturated states, reflecting two distinct states of stellar magnetic activity, both closely linked to the Rossby number $R_{\mathrm{o}}=\frac{2\pi}{\Omega_{\mathrm{crit}}\tau_{\mathrm{cz}}}$. The convective turnover timescale $\tau_{\mathrm{cz}}$ can be mathematically expressed as \citep{2021ApJ...912...65G}:
\begin{equation}
	\tau_{\mathrm{cz}}(r)=\alpha_{\mathrm{MLT}}H_{\mathrm{P}}(r)/v_{\mathrm{c}}(r).
	\label{eq6}
\end{equation}

In Equation (\ref{eq6}), $H_{\mathrm{P}}(r)$ represents the scale height, $v_{\mathrm{c}}(r)$ denotes the convective velocity at radius $r$, and $\alpha_{\mathrm{MLT}}$ stands for the convective mixing length. The turnover timescale of the convective zone, $\tau_{\mathrm{cz}}$, is defined at the location where $r = r_{\mathrm{BCZ}} + 0.5H_{\mathrm{P}}(r)$, with $r_{\mathrm{BCZ}}$ being the radius of the bottom of the outer convection zone. The value of $\alpha_{\mathrm{MLT}}$ is fixed at 1.82.
The region where $R_{\mathrm{o}} \leqslant R_{\mathrm{osat}}$ is termed the saturation region, and vice versa. The parameter $\chi = R_{\mathrm{o}}/R_{\mathrm{osat}}$ depends on the critical value $R_{\mathrm{osat}} = 0.14$ \citep{2018MNRAS.479.2351W}. The solar $R_{\mathrm{o}}$ is approximately 2 \citep{2016MNRAS.462.4442S}, thus we set $\chi = 14$. The constant $m$ is assigned a value of 0.22 \citep{2019A&A...631A..77A}. To reproduce the current rotation period of the Sun, the values of $K$ and $p$ are chosen as $1.2 \times 10^{30}$ $erg$ and 2.6, respectively.
Next, we present the relationship between the orbital angular momentum of the star-planet system and the evolution of the stellar spin angular momentum over time as follows:
\begin{equation}
	\frac{\mathrm{d}L_{\mathrm{orb}}}{\mathrm{d}t}=\frac{1}{2}\frac{M_{*}M_{\mathrm{pl}}}{M_{*}+M_{\mathrm{pl}}}\mathrm{n}a\frac{\mathrm{d}a}{\mathrm{dt}},
	\label{eq7}
\end{equation}
\begin{equation}
	\frac{\mathrm{d}L_{*}}{\mathrm{d}t}=I_* \frac{\mathrm{d}\Omega}{\mathrm{d}t} + \Omega\frac{\mathrm{d}{I_*}}{\mathrm{d}t}=\frac{\mathrm{d}L_{\mathrm{wind}}}{\mathrm{d}t}-\frac{\mathrm{d}L_{\mathrm{orb}}}{\mathrm{d}t}.
	\label{eq8}
\end{equation}
Here, $\mathrm{n}$ represents the orbital angular velocity, a denotes the semi-major axis of the orbit, and $I_*$ stands for the moment of inertia. Since the moment of inertia changes minimally during the main sequence for solar-like stars, we assume $\frac{\mathrm{d}{I_*}}{\mathrm{d}t}=0$.

The rate of change of the planet's orbit over time, we can use the equation (1) of \citet{2012ApJ...751...96P}, the equation (1) was obtained through \citep{1963MNRAS.126..257G,1968aitp.book.....K,2008ApJ...678.1396J}:
\begin{equation}
	\frac{\mathrm{d}a}{\mathrm{d}t} = \frac{9}{2} \sqrt{\frac{G}{aM_*}} (\frac{R_*}{a})^5 \frac{M_\mathrm{pl}}{Q'_*}\left(\frac{{\mathrm{n}}-{\Omega}}{|{\mathrm{n}}-{\Omega}|}\right).
	\label{eq9}
\end{equation}
Here, $Q'_*$ and n represent the tidal quality factor of the star and the orbital frequency of the planet, respectively.
Combining Equations (\ref{eq7}), (\ref{eq8}), and (\ref{eq9}), we obtain the expression of the time derivative of the stellar angular rotation rate:
\begin{equation}
	\frac{\mathrm{d{\Omega}}}{\mathrm{d{t}}} = \frac{1}{I_*}\left(\frac{9}{4{Q'_{_*}}}\frac{{M_*}^\frac{1}{2}{M_{\mathrm{pl}}}^2{R_*}^5}{G(M_*+M_{\mathrm{pl}})^\frac{5}{2}}\left(\frac{{\mathrm{n}}-{\Omega}}{|{\mathrm{n}}-{\Omega}|}\right){\mathrm{n}}^4 - T_{\mathrm{0}}(\frac{\tau_{\mathrm{cz}}}{\tau_{\mathrm{cz\odot}}})^p(\frac{\Omega}{\Omega_\odot})^{p+1}\right)        
	\label{eq10}
	(unsaturated),
\end{equation}

\begin{equation}
	\frac{\mathrm{d{\Omega}}}{\mathrm{d{t}}} = \frac{1}{I_*}\left(\frac{9}{4{Q'_{_*}}}\frac{{M_*}^\frac{1}{2}{M_{\mathrm{pl}}}^2{R_*}^5}{G(M_*+M_{\mathrm{pl}})^\frac{5}{2}}\left(\frac{{\mathrm{n}}-{\Omega}}{|{\mathrm{n}}-{\Omega}|}\right){\mathrm{n}}^4 - T_{\mathrm{0}}{\chi}^p(\frac{\Omega}{\Omega_\odot})\right)      (saturated).
	\label{eq11}
\end{equation}

\subsection{Orbital Stability - Long-term Double Synchronization} \label{sec:style}

\citet{2015A&A...574A..39D} first considered the orbital stability of gaseous giant planet systems in the presence of stellar angular momentum loss. However, due to the existence of stellar magnetic braking, this stability may be disrupted as the system evolves, leading to a state known as pseudo-stability of the planetary system. The total angular momentum $L$, critical angular momentum $L_{c_0}$, and critical orbital rate $n_{c_0}$ of the system are defined as follows:
\begin{equation}
	L = G^{2/3}\frac{M_{*}M_{\mathrm{pl}}}{(M_{*}+M_{\mathrm{pl}})^{1/3}}n^{-1/3}+(\frac{\Omega}{n}I_{*}+I_{\mathrm{sp}})\mathrm{n},
	\label{eq12}
\end{equation}

\begin{equation}
	L_{c_0}=4\left[\frac{G^2}{3^3}\frac{M_{*}^3M_\mathrm{pl}^3}{M_{*}+M_\mathrm{pl}}(I_{*}+I_{\mathrm{sp}})\right]^{1/4},
	\label{eq13}
\end{equation}

\begin{equation}
	n_{c_0}=\left[\frac{G^2}{3^3}\frac{M_{*}^3M_\mathrm{pl}^3}{M_{*}+M_\mathrm{pl}}\right]^{1/4}(I_{*}+I_{\mathrm{sp}})^{-3/4}.
	\label{eq14}
\end{equation}
where $I_*$ and $I_{\mathrm{sp}}$ are the stellar and planetary moment of inertia. Therefore, as long as the mean motion of the orbit is $n < \left(\frac{3}{4}\right)^{3/4}n_{\mathrm{c_0}}$ and $L > L_{\mathrm{c_0}}$, the system can achieve a stable equilibrium.

Long-term double synchronization is a special case of system stability, which requires the introduction of a physical quantity known as the equilibrium stellar rotation rate $\Omega_\mathrm{sta}$. We solved for the equilibrium stellar rotation rate $\Omega_\mathrm{sta}$ under the balance of torque due to wind and tidal forces. We found that long-term double synchronization can be triggered and maintained for an extended period when the stellar rotation rate is approximately equal to the planetary orbital angular velocity and the evolutionary trajectory of $\Omega_\mathrm{sta}$ aligns closely with the trajectory of $\Omega_\mathrm{sta}=n$. This mechanism significantly slows down the orbital decay rate until the stable conditions are disrupted, leading to rapid planet engulfment by the star. When $\frac{d{\Omega}}{d{t}}=0$ represents the equilibrium stellar angular frequency $\Omega=\Omega_\mathrm{sta}$. \citep{2024MNRAS.529.2893G}

\begin{table}
	
	\caption{The parameters adopted in Figure \ref{fig:type}\label{tab:Parameter}}
	\centering
	\begin{tabular}{|l|c|c|c|c|c|c|}
		\hline
		migration states & $M_{*}$ ($M_{\odot}$) & $\mathrm{\mathrm{\mathrm{[Fe/H]}}}$ (dex) & $M_\mathrm{pl}$ ($M_\mathrm{J}$) & $P_\mathrm{rot, ini}$ (days) & $Q_{*}$ & $a_\mathrm{ini}$ (au) \\
		\hline
		OUTG & 1.1 & -0.5 & 1.0 & 3.0 & $10^{5}$ & 0.05 \\
		OUTS & 1.1 & -0.5 & 1.0 & 3.0 & $10^{6}$ & 0.05 \\
		IN & 1.1 & -0.5 & 1.0 & 8.0 & $10^{6}$ & 0.05 \\
		OUTG DS & 1.1 & -0.5 & 8.0 & 3.0 & $10^{6}$ & 0.06 \\
		OUTS DS & 1.1 & -0.5 & 4.0 & 3.0 & $10^{6}$ & 0.05 \\
		IN DS & 1.1 & -0.5 & 4.0 & 8.0 & $10^{6}$ & 0.06 \\
		\hline
	\end{tabular}
\end{table}

\section{Data Sets} \label{sec:data}

Our star-planet system model consists of six initial input parameters as follows: stellar mass $M_* (0.8$\,$M_{\odot} < M_* < 1.3$\,$M_{\odot})$, planetary mass $M_\mathrm{pl} (0.1$\,$M_\mathrm{J}< M_\mathrm{pl} < 8.0$\,$M_\mathrm{J})$, initial stellar spin rate $\omega_{\mathrm{rot,ini}} (1.57$\,$\Omega_{\odot} < \omega_{\mathrm{rot,ini}} < 17.5$\,$\Omega_{\odot})$, initial semi-major axis of the planet's orbit $a (0.02$\,$au[4.31$\,$R_*] < a < 0.08$\,$au[17.24$\,$R_*])$, tidal quality factor $Q'_* (10^{5} < Q'_* < 10^{9})$, and stellar metallicity $\mathrm{\mathrm{[Fe/H]}} (-0.5$\,$dex < \mathrm{\mathrm{[Fe/H]}} < +0.5$\,$dex)$. Our work mainly focuses on star-planet systems, and we generated 15,745 evolutionary curves by randomly sampling points within the parameter space using the stellar-planet interaction model. We discuss the results of the model evolution in two parts.

\begin{figure*}
	\includegraphics[width=0.95\textwidth]{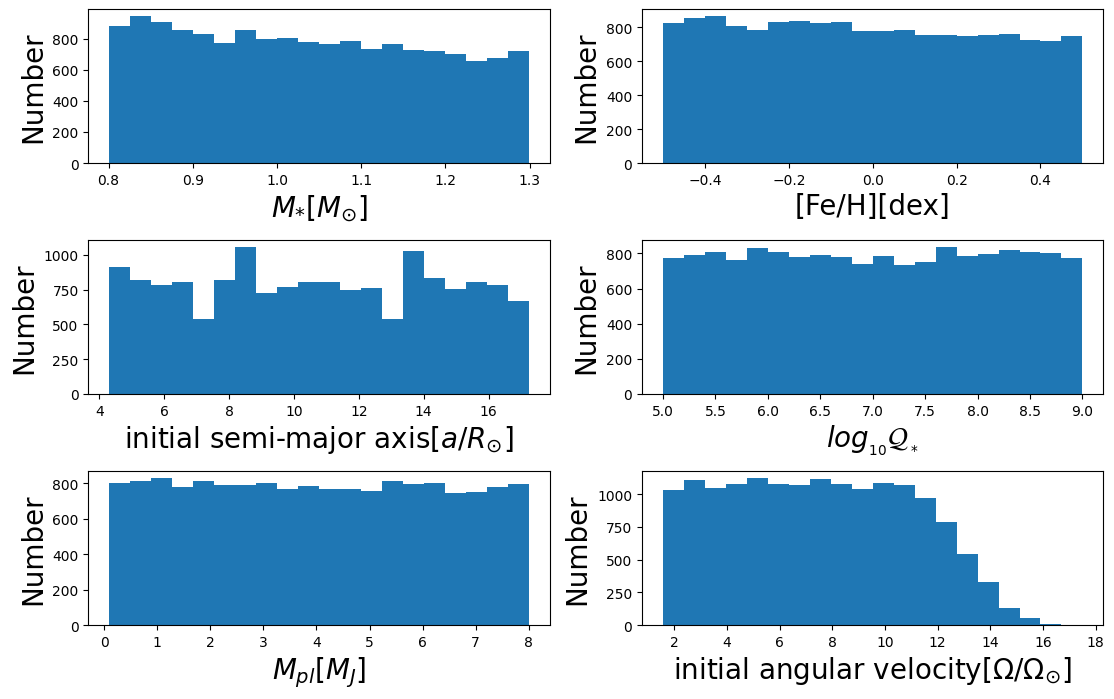}
	\caption{The initial input parameter distributions are shown in the figure, where the initial semi-major axis of the planet is expressed in units of solar radii, and the initial stellar rotation frequency is expressed in units of solar rotation frequency.   \label{fig:histogram}}
\end{figure*}

In the first part, we applied artificial neural networks to the evolutionary curves generated by the stellar-planet interaction model for regression training using the six initial parameters as inputs. In the second part, we categorized the evolutionary curves into five classes based on the survival status of the planet and the type of evolution. We used lightGBM to perform binary classification on the types of evolutionary curves, including determining whether a curve belongs to the category of long-term double synchronization, given any six initial parameter inputs. Thus, our work can be divided into regression and classification. Figure \ref{fig:histogram} shows the distribution of the six initial input parameters in the 15,745 evolutionary curves. Overall, the distribution of the initial parameters in our dataset is relatively uniform. Regarding the initial stellar spin frequency, we observed a sharp decrease in data with spin rates exceeding 11 times the solar rotation frequency. This is because in our model, stars with larger initial radii in the main sequence phase are less likely to reach their critical rotation frequency when they start with excessively fast initial spin rates, making it difficult to generate curves in these cases.

\subsection{Regression Data} \label{subsec:style}
We consider four evolving quantities as outputs in our model: stellar effective temperature ($logT_\mathrm{eff}$), stellar radius ($R_*$), planetary orbital period ($P_{\mathrm{orb}}$), and stellar rotation period ($P_{\mathrm{rot}}$). Here, we define the maximum age of a star-planet system. We determine the maximum age when any of the following conditions are met: ($\romannumeral1$) The end of the main sequence phase for the star, ($\romannumeral2$) The semi-major axis of the planet's orbit reaches the Roche limit (1.44$\,$$R_*$), \citet{1999ssd..book.....M} proposed that the Roche limit varies between 1.44 and 2.44 $R_*$. In this study, we adopt a conservative assumption based on the work of \citet{2020A&A...643A..34O}, where the Roche limit is set at 1.44 $R_*$. This means that angular momentum is no longer transferred to the star once the planet is disrupted. ($\romannumeral3$) The system age reaches 14.0 Gyr. For each of the four evolving quantities of any given sample, we perform linear interpolation with 500 points using the interp1d function in Python between the initial age and the maximum age to ensure consistent data points for each sample.

During the training process, we observed suboptimal performance of our neural network model in regressing stellar effective temperature ($\log T_\mathrm{eff}$) and stellar radius ($R_*$). This is attributed to the model's assumption that interactions between planets and stars do not influence variations in these stellar properties. Moreover, when the distance between a planet and its host star reaches the Roche limit, the model halts, resulting in the termination of stellar evolution. Consequently, data for $\log T_\mathrm{eff}$ and $R_*$ during the main sequence phase exhibit incompleteness. To address this issue, we curated a separate dataset comprising stellar evolution curves for stars without planets. To enhance the regression capability of the neural network, we needed to appropriately handle data incompleteness. By randomly sampling points in parameter space, we generated 7500 evolutionary curves spanning the range of stellar mass and metallicity. The distributions of stellar mass and metallicity align with those depicted in Figure \ref{fig:histogram}. In this context, we defined the maximum age of the main sequence for stars based on the model's termination point. The treatment of stellar samples paralleled that of planetary system samples. Utilizing stellar mass and metallicity as parameters, we predicted stellar effective temperature ($\log T_\mathrm{eff}$) and stellar radius ($R_*$) for individual stars in this dataset.

\subsection{Classified Data} \label{subsec:style}
For categorical data, we also include the six initial input parameters as in the regression data, but here our focus is on the orbital evolution of planets in high-dimensional space. For each curve, we categorize them into six classes based on the planetary migration states, namely: 1. planets migrating outward and reaching maximum age with semimajor axis greater than the initial value (OUTG), 2. During the early stages of evolution, the planet experiences an outward migration phase due to the stellar rotation rate exceeding the orbital velocity of the planet (The phase is much shorter compared to "OUTG"). Subsequently, as the stellar rotation rate becomes slower than the orbital velocity of the planet, the planet undergoes inward migration (The phase is much longer compared to "OUTG"). Upon reaching its maximum age, the semi-major axis of the planet is smaller than its initial value (OUTS), 3. planets migrating inward only (IN), 4. Type 1 and experiencing long-term double synchronization (OUTG DS), 5. Type 2 and experiencing long-term double synchronization (OUTS DS), and 6. Type 3 and experiencing long-term double synchronization (IN DS).

\begin{figure*}
	\includegraphics[width=0.95\textwidth]{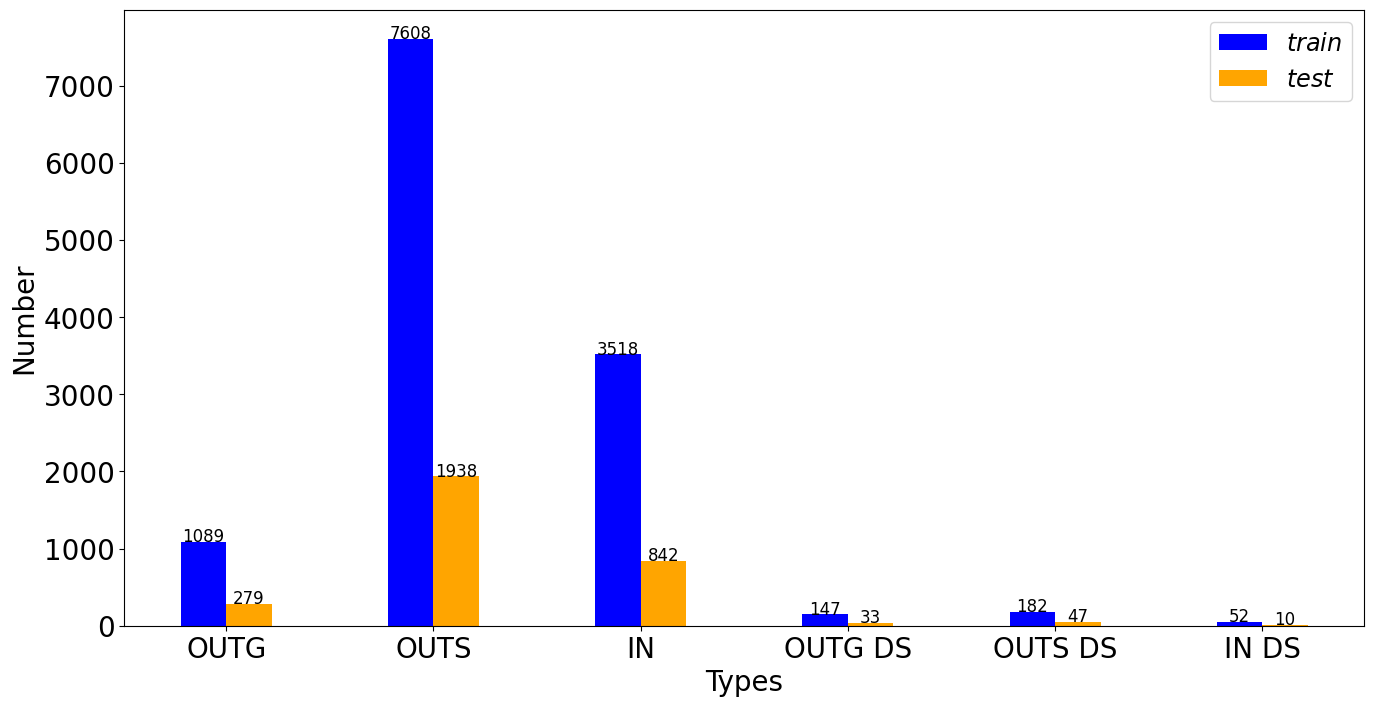}
	\caption{The number of samples in 6 subclass of (a) OUTG (b) OUTS (c) IN (d) OUTG DS (e) OUTS DS and (f) IN DS in training set and test set. The theoretical model calculates that only 3\% of the samples exhibit double synchronization. Meanwhile, in Figure 6(a) of \citep{2024MNRAS.529.2893G}, it can be find that there are approximately 10 observed systems near  $\Omega$/n=1. This indicates that the number of samples in double synchronization state is very low.   \label{fig:CN}}
\end{figure*}

\begin{figure*}
	\centering
	\begin{subfigure}[t]{0.3\textwidth}
		\centering
		\includegraphics[width=\textwidth]{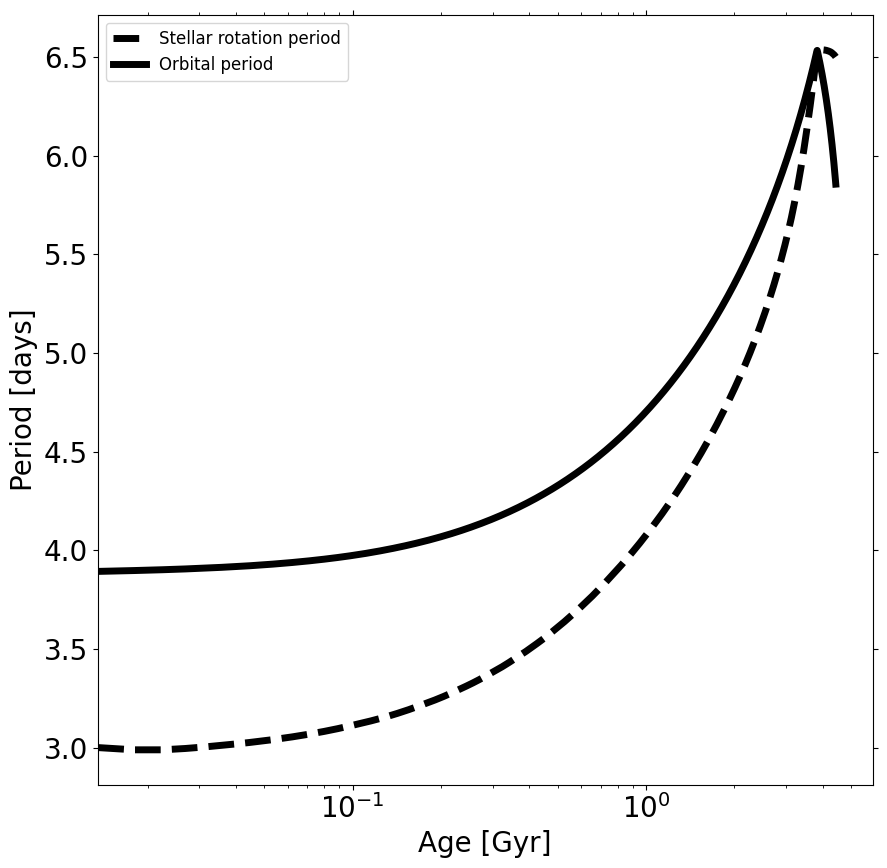}
		\caption{}
		\label{fig:type:a}
	\end{subfigure}
	\begin{subfigure}[t]{0.3\textwidth}
		\centering
		\includegraphics[width=\textwidth]{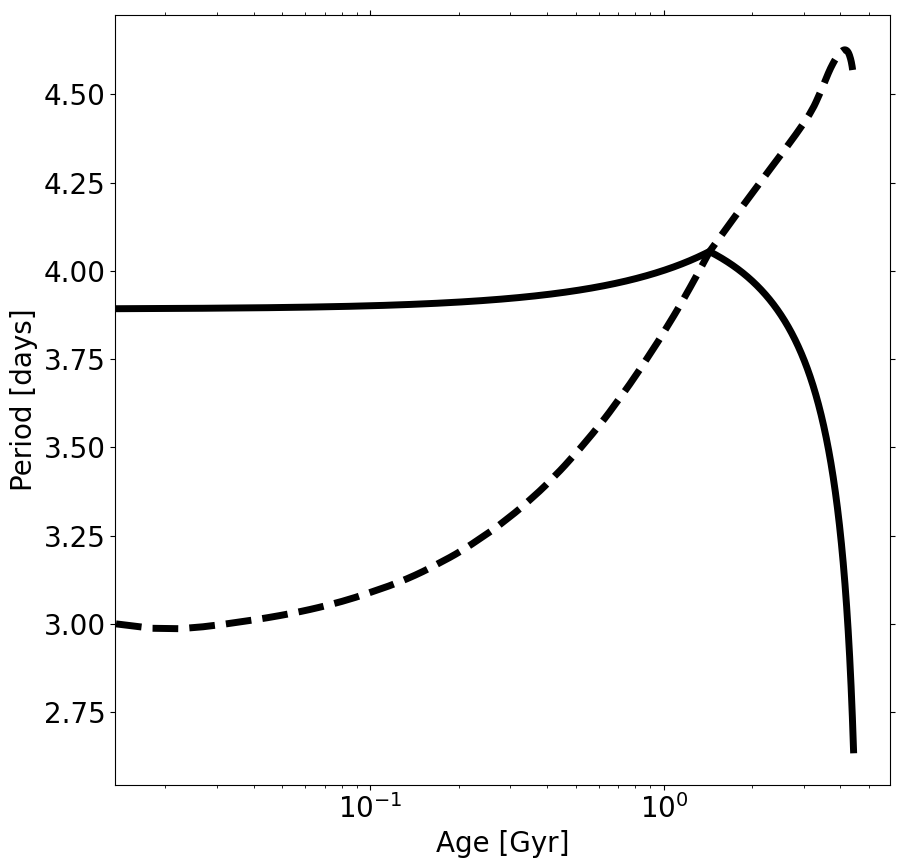}
		\caption{}
		\label{fig:type:b}
	\end{subfigure}
	\begin{subfigure}[t]{0.3\textwidth}
		\centering
		\includegraphics[width=\textwidth]{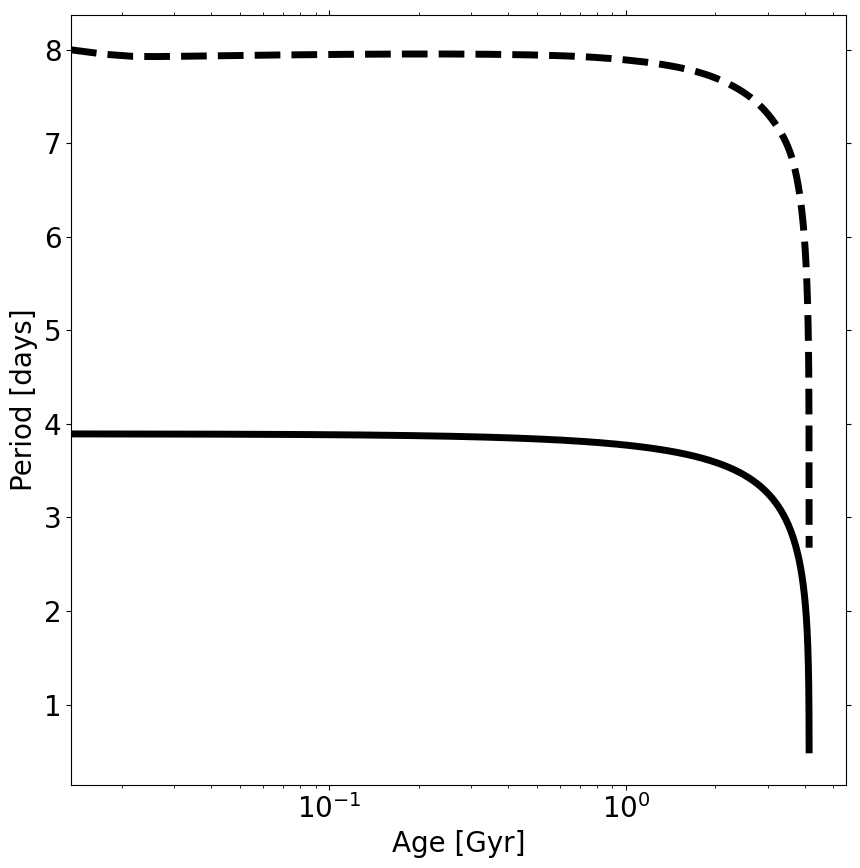}
		\caption{}
		\label{fig:type:c}
	\end{subfigure}
	\begin{subfigure}[t]{0.3\textwidth}
		\centering
		\includegraphics[width=\textwidth]{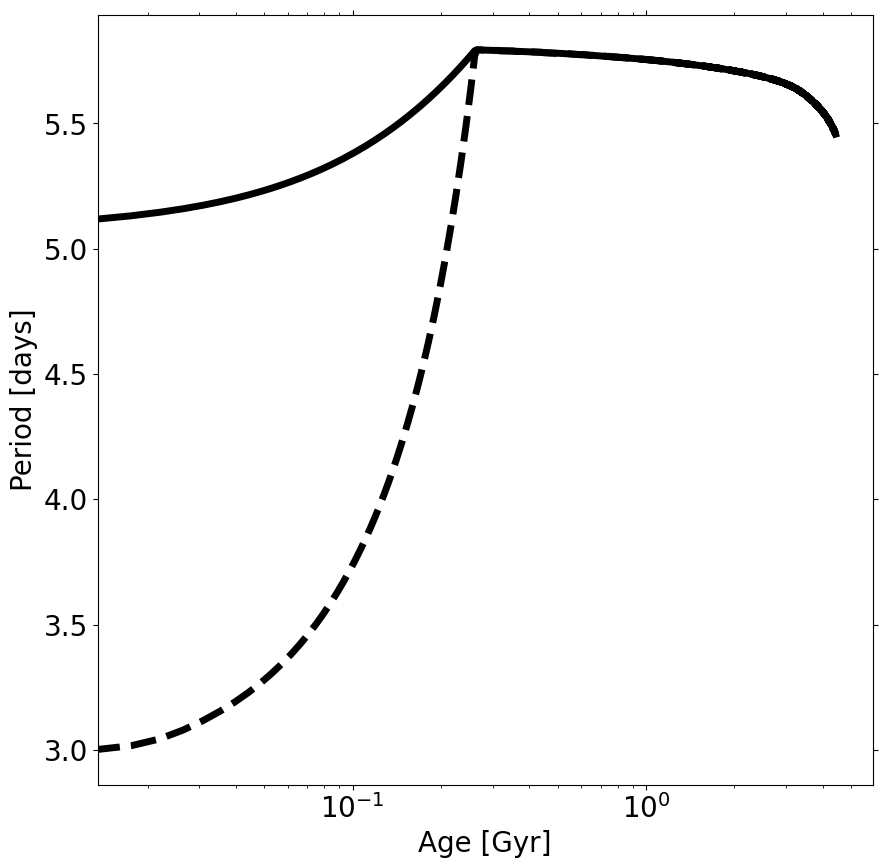}
		\caption{}
		\label{fig:type:d}
	\end{subfigure}
	\begin{subfigure}[t]{0.3\textwidth}
		\centering
		\includegraphics[width=\textwidth]{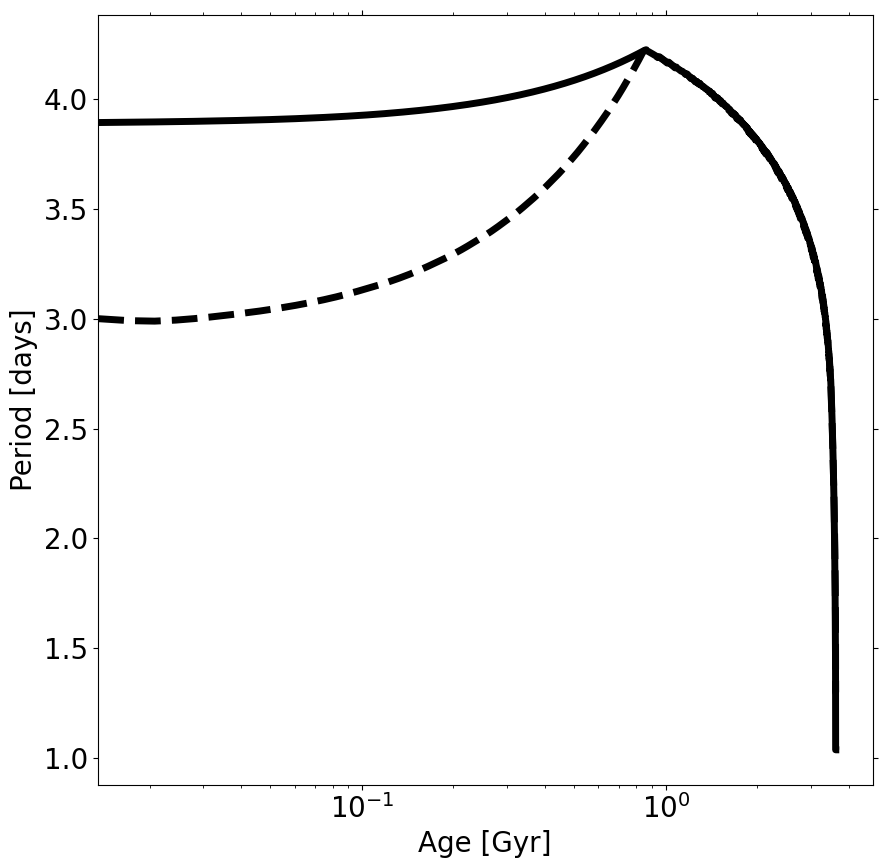}
		\caption{}
		\label{fig:type:e}
	\end{subfigure}
	\begin{subfigure}[t]{0.3\textwidth}
		\centering
		\includegraphics[width=\textwidth]{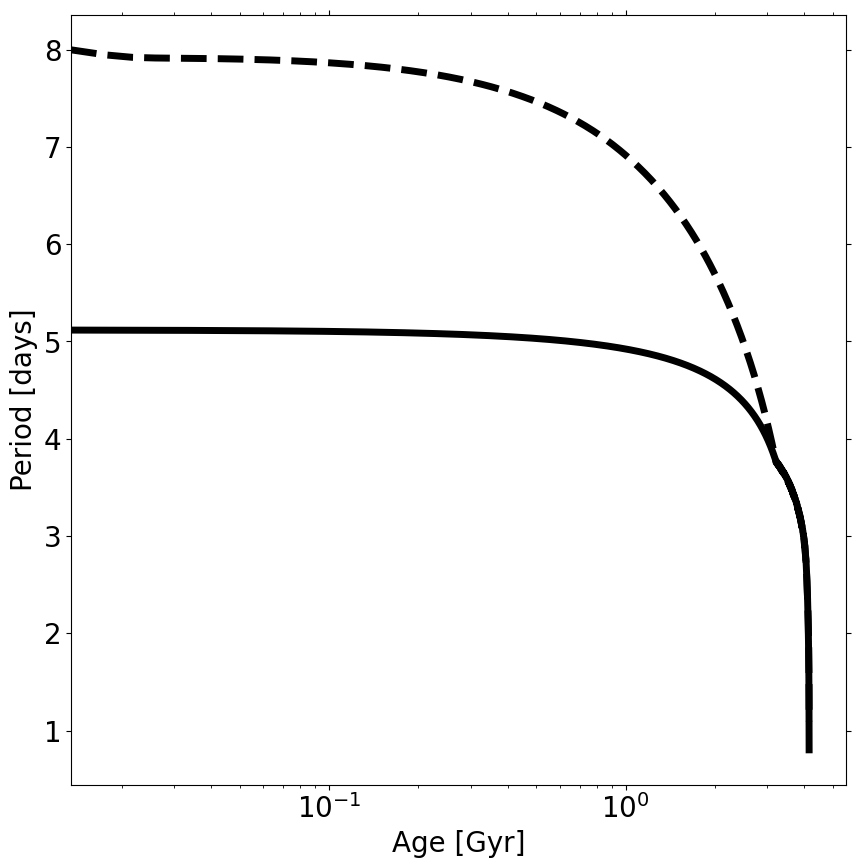}
		\caption{}
		\label{fig:type:f}
	\end{subfigure}
	\caption{The figure depicts six states of stellar migration: a. OUTG, b. OUTS, c. IN, d. OUTG DS, e. OUTS DS and f. IN DS.
		\label{fig:type}}
\end{figure*}

In Figure \ref{fig:CN}, we also present the statistics of the number of training and test sets for each type among the 15,745 star-planet system samples. Additionally, for each of the six types, we showcase an example and illustrate the age-period relation in Figure \ref{fig:type}(In order to provide a clearer illustration of these six migration states, we consider samples with uniform stellar mass of 1.1 $M_{\odot}$ and metallicity of -0.5 dex (where the stellar convective envelope is thin, making long-term double synchronization more likely). In our sample, the stellar rotation period can reach up to 80 days during evolution, while the planetary orbital period is less than 10 days. To showcase double synchronization samples, we select samples throughout the evolutionary stage where the stellar rotation period is less than 10 days). The initial input parameters for each example are provided in Table \ref{tab:Parameter}.

\section{Models and methods} \label{sec:MAM}

\subsection{Establishment and parameter setting of neural network model} \label{subsec:style}
We employed an artificial neural network (NN), also known as a multilayer perceptron (MLP), which consists of an input layer, hidden layers, and an output layer. In our work, for the four evolution parameters, the neural network comprises one input layer, four hidden layers with 1000 neurons each, and one output layer. For the maximum system age and main-sequence maximum age, the neural network consists of one input layer, two hidden layers with 20 neurons each, and one output layer.
As mentioned in the previous section, the input layer for the stellar model includes 2 input parameters, while the stellar-planet model includes 6 input parameters to train and generate the evolution parameters and age. Both the output layer and hidden layers have activation functions, which introduce nonlinearity, allowing the network to approximate any nonlinear function and tackle nonlinear problems. We chose the ReLU function as the activation function.
In our model, the loss function represents the difference between the actual and predicted values. During the training process, the change in the loss function is used to select a better model. For the stellar effective temperature and stellar radius, the mean absolute error (MAE) is used as the loss function, which converges faster. For the stellar rotation period, planetary orbital period, maximum age, and maximum main-sequence age, the mean squared error (MSE) is used as the loss function for faster convergence.

Here are the mathematical expressions for MSE (Mean Squared Error) and MAE (Mean Absolute Error). In these expressions, $n$ represents the number of samples, $y_i$ represents the true values, and $\hat{y}_i$ represents the predicted values by the model.

\begin{equation}
	MSE = \frac{1}{n} \sum_{i=1}^{n} (y_i - \hat{y}_i)^2
\end{equation}

\begin{equation}
	MAE = \frac{1}{n} \sum_{i=1}^{n} |y_i - \hat{y}_i|
\end{equation}

We randomly split our samples into an 80$\%$ training set and a 20$\%$ testing set.

\subsection{Selection and Evaluation of Classification Models} \label{subsec:style}
In this section, our model is a binary classifier using the gradient boosting decision tree (GBDT) model from the popular machine learning framework lightGBM \citep{ke2017advances}. Specifically, we use the boosting='gbdt' parameter. The GBDT model is based on the idea of residual fitting, where each tree is trained to fit the residuals of the previous trees, progressively improving the model performance. The training objective of each tree is to minimize the negative gradient of the loss function (see \citep{friedman2001annals}). We set the learning rate as learning$\_$rate=0.1, the number of leaves per tree as num$\_$leaves=31, and the maximum tree depth as max$\_$depth:=-1 (indicating no depth limit).We chose objective='multiclass' to address the multi-class classification problem.

We observe severe class imbalance in some categories, especially for the "long-term double synchronization" case where the positive examples account for less than 2$\%$. Since our data is limited and there are significant differences between positive and negative samples in certain categories, we address this issue in two ways. Firstly, we employ the Synthetic Minority Over-sampling Technique (SMOTE) from the imblearn. over$\_$sampling module in Python to oversample the extremely imbalanced samples, thereby improving the distribution of sample categories. However, we only oversample the training set and augment the minority samples to a total of 1500 samples to enhance our training model. Secondly, we employ K-fold stratified cross-validation (with K=10) for each category to assess the robustness of the model, following the method proposed by \citet{picard1984cross}. Specifically, we randomly divide the training set into 10 non-overlapping subsets, ensuring an equal proportion of positive and negative samples in each subset. Then, we train on 9 subsets and test on the remaining subset, repeating this process 10 times to create 10 different combinations. Finally, we employ the VotingClassifier with voting='hard' setting, where each model gives a discrete prediction, and the final prediction is determined by majority voting. By conducting stratified 10-fold cross-validation and model voting, we improve the stability of the model, reduce the risk of overfitting, enhance model performance, and leverage the advantages of diverse models to obtain a more reliable and accurate optimal model.

\section{Result} \label{sec:result}

\subsection{Results and Error Analysis of Evolutionary Quantities} \label{subsec:style1}
\begin{figure*}
	\centering
	\begin{subfigure}[t]{0.3\textwidth}
		\centering
		\includegraphics[width=\textwidth]{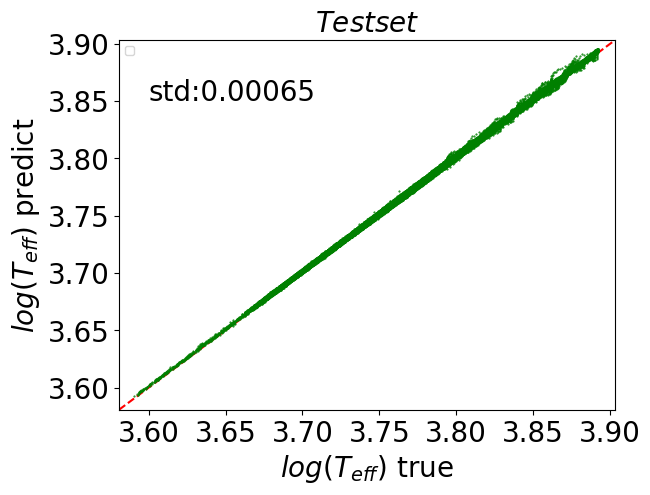}
		\caption{}
		\label{fig:regression1:a}
	\end{subfigure}
	\begin{subfigure}[t]{0.3\textwidth}
		\centering
		\includegraphics[width=\textwidth]{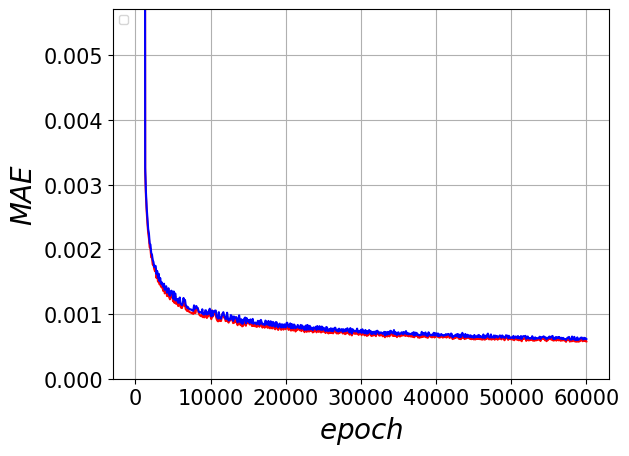}
		\caption{}
		\label{fig:regression1:b}
	\end{subfigure}
	\begin{subfigure}[t]{0.3\textwidth}
		\centering
		\includegraphics[width=\textwidth]{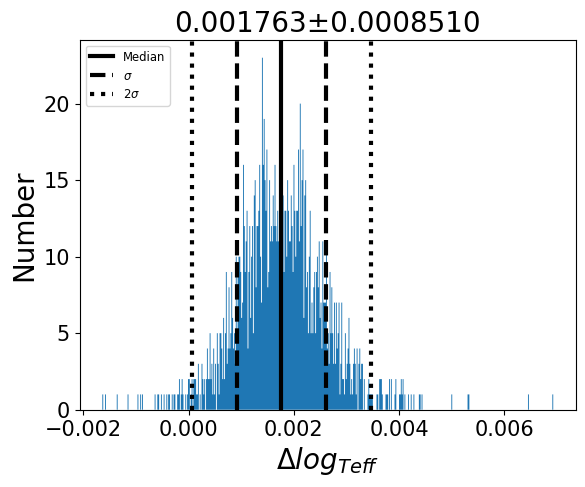}
		\caption{}
		\label{fig:regression1:c}
	\end{subfigure}
	\begin{subfigure}[t]{0.3\textwidth}
		\centering
		\includegraphics[width=\textwidth]{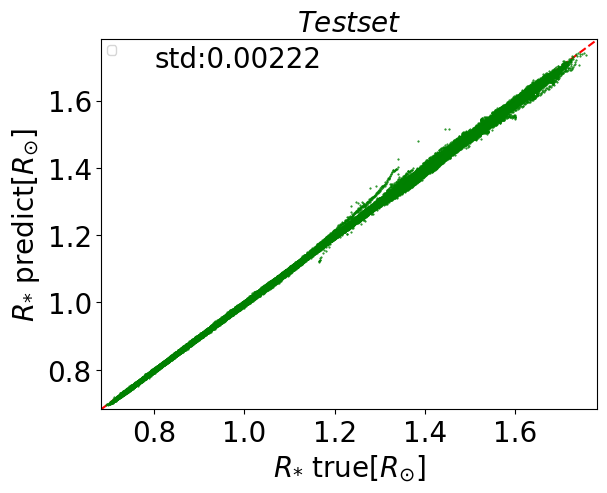}
		\caption{}
		\label{fig:regression1:d}
	\end{subfigure}
	\begin{subfigure}[t]{0.3\textwidth}
		\centering
		\includegraphics[width=\textwidth]{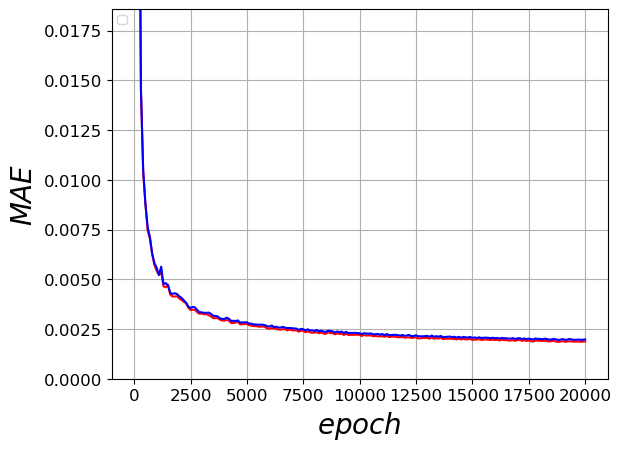}
		\caption{}
		\label{fig:regression1:e}
	\end{subfigure}
	\begin{subfigure}[t]{0.3\textwidth}
		\centering
		\includegraphics[width=\textwidth]{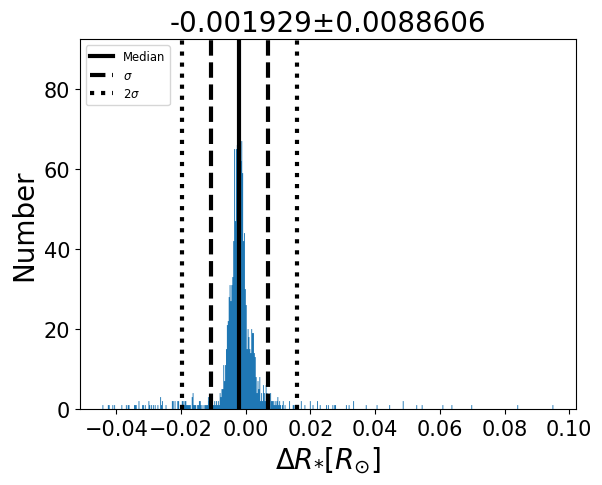}
		\caption{}
		\label{fig:regression1:f}
	\end{subfigure}
	\begin{subfigure}[t]{0.3\textwidth}
		\centering
		\includegraphics[width=\textwidth]{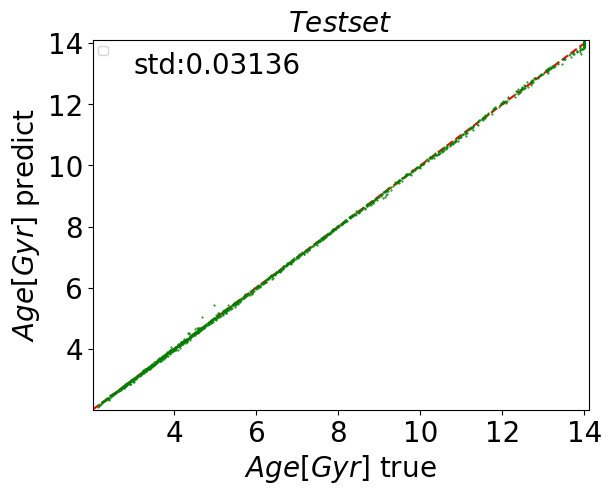}
		\caption{}
		\label{fig:regression1:g}
	\end{subfigure}
	\begin{subfigure}[t]{0.3\textwidth}
		\centering
		\includegraphics[width=\textwidth]{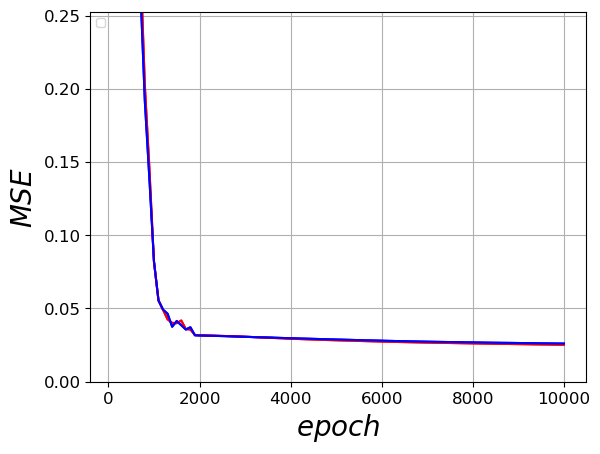}
		\caption{}
		\label{fig:regression1:h}
	\end{subfigure}
	\begin{subfigure}[t]{0.3\textwidth}
		\centering
		\includegraphics[width=\textwidth]{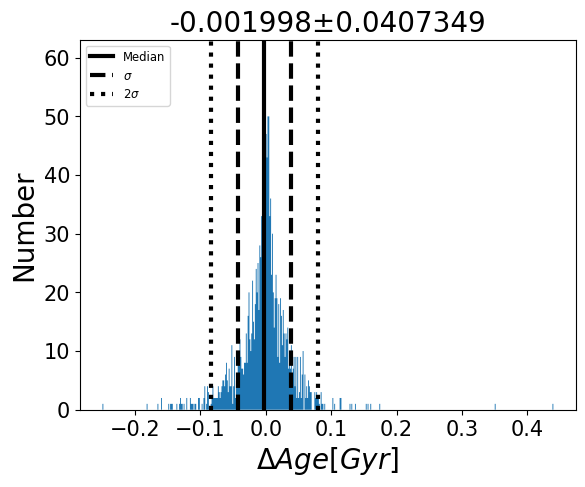}
		\caption{}
		\label{fig:regression1:i}
	\end{subfigure}
	\caption{The figure illustrates the results of using NN for regression on stellar samples, including stellar effective temperature, stellar radius, and maximum main sequence age. Left Panel: the green dots represent the predicted values using the neural network model on the test set, the red dashed line represents the 45° line where predicted values are equal to true values. Middle Panel: The learning curve depicting the change in MAE (MSE) for the training and test sets during the learning process (recorded every 100 epochs, where the red and blue curves represent the training and test sets, respectively). Right Panel: the error distribution histogram of the predicted values and true values for the training set. The mean ($\mu$) and variance ($\sigma$) of the distribution are provided above each plot, and the black vertical lines indicate the percentiles (2nd, 16th, 50th, 84th, and 98th) of the samples. The top plot corresponds to stellar effective temperature, the middle plot represents stellar radius, and the bottom plot shows the maximum main sequence age of the stars.\label{fig:regression1}}
\end{figure*}

\begin{figure*}
	\centering
	\begin{subfigure}[t]{0.3\textwidth}
		\centering
		\includegraphics[width=\textwidth]{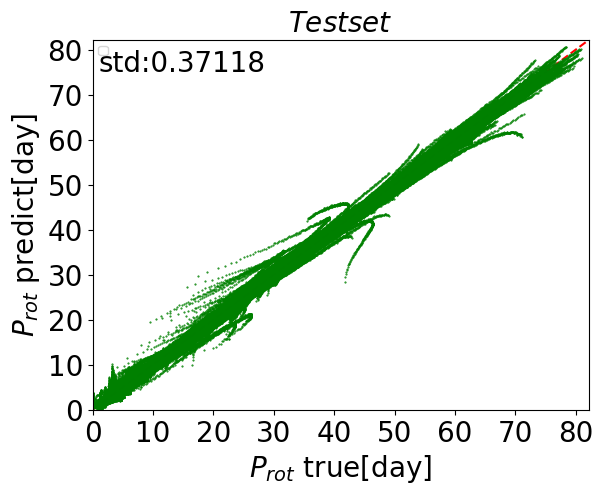}
		\caption{}
		\label{fig:regression2:a}
	\end{subfigure}
	\begin{subfigure}[t]{0.3\textwidth}
		\centering
		\includegraphics[width=\textwidth]{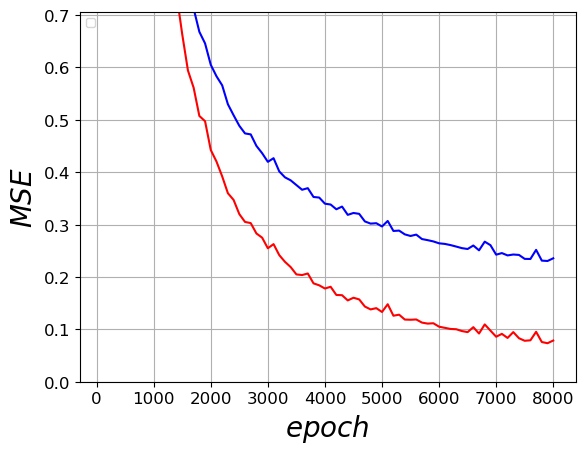}
		\caption{}
		\label{fig:regression2:b}
	\end{subfigure}
	\begin{subfigure}[t]{0.3\textwidth}
		\centering
		\includegraphics[width=\textwidth]{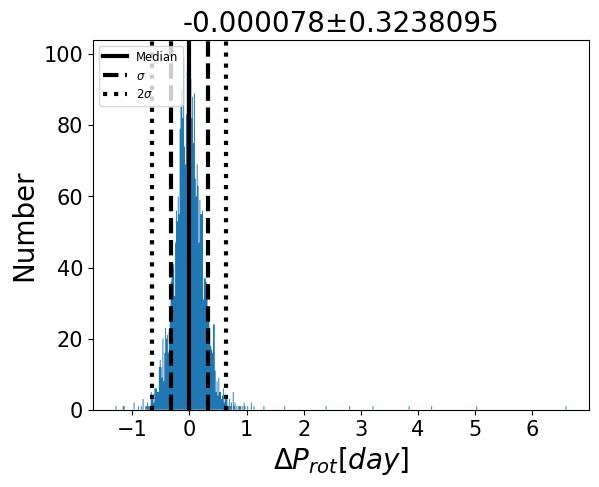}
		\caption{}
		\label{fig:regression2:c}
	\end{subfigure}
	\begin{subfigure}[t]{0.3\textwidth}
		\centering
		\includegraphics[width=\textwidth]{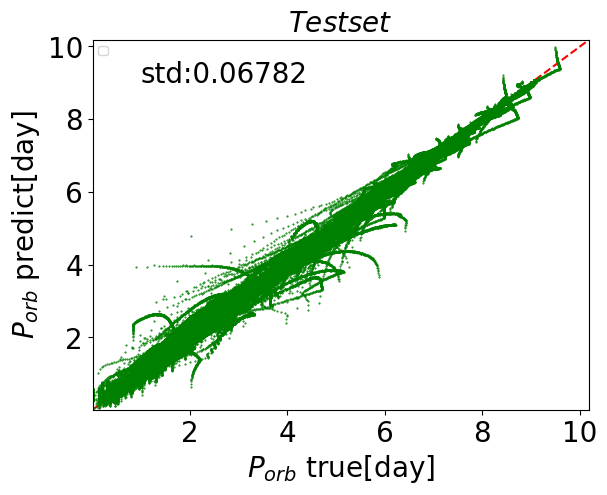}
		\caption{}
		\label{fig:regression2:d}
	\end{subfigure}
	\begin{subfigure}[t]{0.3\textwidth}
		\centering
		\includegraphics[width=\textwidth]{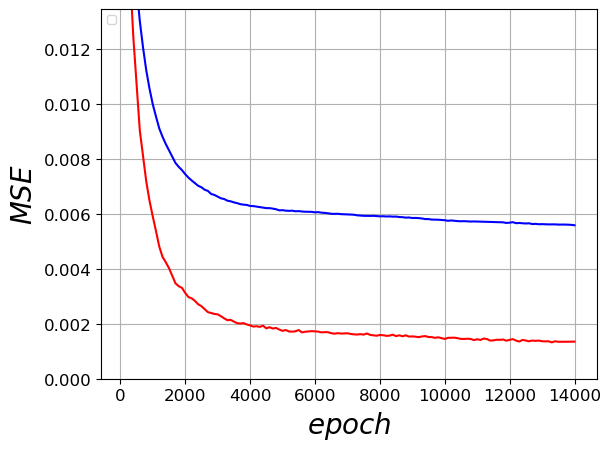}
		\caption{}
		\label{fig:regression2:e}
	\end{subfigure}
	\begin{subfigure}[t]{0.3\textwidth}
		\centering
		\includegraphics[width=\textwidth]{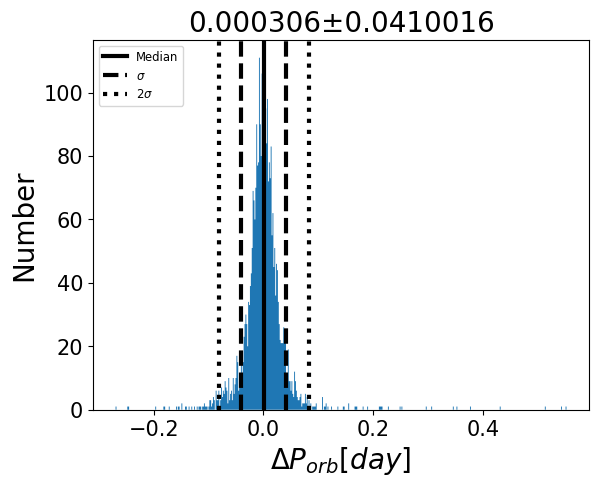}
		\caption{}
		\label{fig:regression2:f}
	\end{subfigure}
		\begin{subfigure}[t]{0.3\textwidth}
		\centering
		\includegraphics[width=\textwidth]{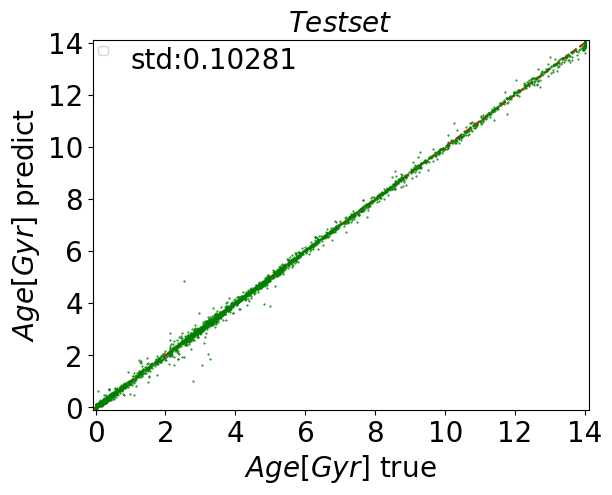}
		\caption{}
		\label{fig:regression2:g}
	\end{subfigure}
	\begin{subfigure}[t]{0.3\textwidth}
		\centering
		\includegraphics[width=\textwidth]{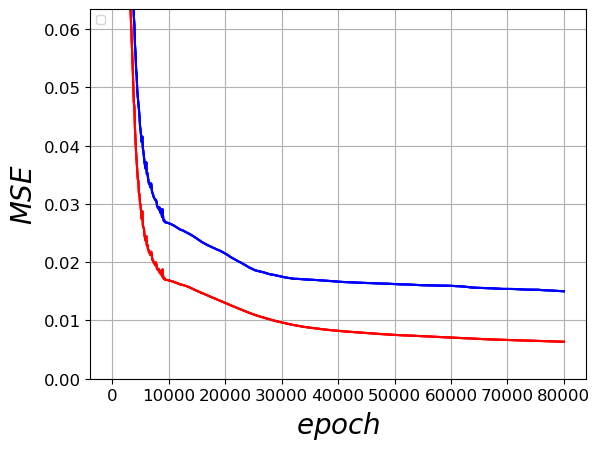}
		\caption{}
		\label{fig:regression2:h}
	\end{subfigure}
	\begin{subfigure}[t]{0.3\textwidth}
		\centering
		\includegraphics[width=\textwidth]{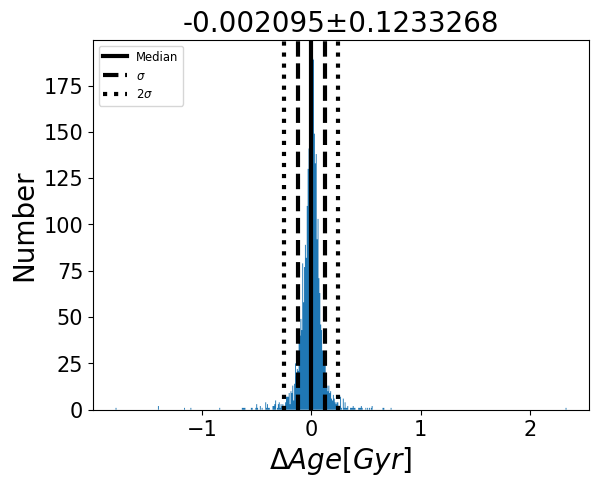}
		\caption{}
		\label{fig:regression2:i}
	\end{subfigure}
	\caption{Similar to Figures \ref{fig:regression1}, we used NN regression on a star-planet system sample to predict the stellar rotation period, planetary orbital period, and maximum system age, with corresponding plots for each variable.
		\label{fig:regression2}}
\end{figure*}

In this section, we present the training results of four evolutionary quantities in regression problems, as well as the maximum age of single stars and star-planet systems. We show the $45^{\circ}$ line of the true values and predicted values of the test set, as well as their residuals in Figures \ref{fig:regression1} and \ref{fig:regression2} (the closer to the $45^{\circ}$ line, the closer the predicted value is to the true value). We also provide a histogram of the residual distribution of the test set, giving the mean $\mu$ and variance $\sigma$ of the distribution. We also provide the range of the 2$\sigma$ distribution in the residual distribution histogram to illustrate the range of our model's error distribution. We also plot the learning curve to describe the changes in MAE (MSE) of the training set and test set during our learning process (recorded every 100 epochs). Residuals refer to the standard deviation of the difference between the predicted value and the true value.
Overall, we found that the regression parameters obtained by training on star samples have a smaller variance than those obtained by training on star-planet system samples. This is because the input parameters of star samples are only two, which makes the evolutionary model simpler compared to the six input parameters of star-planet samples. Additionally, the sample grid is denser, and there are more samples for each average parameter, which makes the neural network model more accurate in regression. We found that the standard deviation of the residual for the effective temperature ($\log T_{\text{eff}}$) in the test set is 0.00065. The distribution of the deviation between predicted and true values is 0.001763 ± 0.0008510. Moreover, over 95$\%$ of the samples exhibit a deviation within the range of $[0.0001, 0.0193]$. For the star radius, the standard deviation of the residual is 0.00222$,R_*$. The difference between predicted and true values is distributed as -0.001929 ± 0.0088606. Again, over 95$\%$ of the samples show a deviation within $[-0.0197, 0.0158]$. Regarding the main sequence maximum age of stars, the standard deviation of the residual is 0.03136 Gyr. The difference between predicted and true values is distributed as -0.001998 ± 0.0407349. Over 95$\%$ of the samples exhibit a deviation within $[-0.0835, 0.0795]$. For the star rotation period, the standard deviation of the residual is 0.37118 days. The difference between predicted and true values is distributed as -0.000078 ± 0.3238095. Again, over 95$\%$ of the samples demonstrate a deviation within $[-0.648, 0.648]$. Concerning the planet's orbital period, the standard deviation of the residual is 0.06782 days. The difference between predicted and true values is distributed as 0.000306 ± 0.0410016. Over 95$\%$ of the samples show a deviation within $[-0.082, 0.082]$. Lastly, for the maximum age of star-planet systems, the residual standard deviation is 0.10281 Gyr. The difference between predicted and true values is distributed as -0.002095 ± 0.1233268. Over 95$\%$ of the samples have a deviation within $[-0.249, 0.245]$. Overall, our regression model has high accuracy. For any given star-planet system within our parameter space, we can use six initial input parameters to quickly generate a theoretical curve. Each evolutionary curve consists of 500 points, and set our disk locking time to 0.013$\,$Gyr. By dividing the difference between the maximum age and the initial disk time stamp by 500, we can obtain the age corresponding to each point, which allows us to obtain the values of the four evolutionary quantities at any given time for any star-planet system.

\subsection{Classification Results and Error Analysis} \label{subsec:style2}
\begin{figure*}
	\centering
	\begin{subfigure}[t]{0.3\textwidth}
		\centering
		\includegraphics[width=\textwidth]{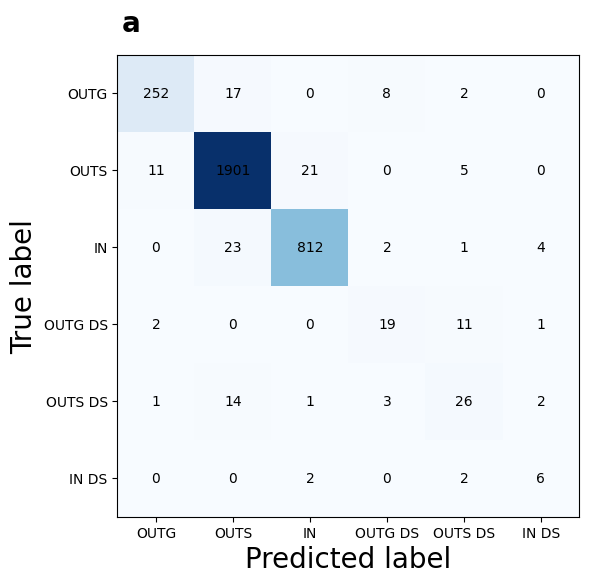}
		\caption{}
		\label{fig:lgb6:a}
	\end{subfigure}
	\begin{subfigure}[t]{0.3\textwidth}
		\centering
		\includegraphics[width=\textwidth]{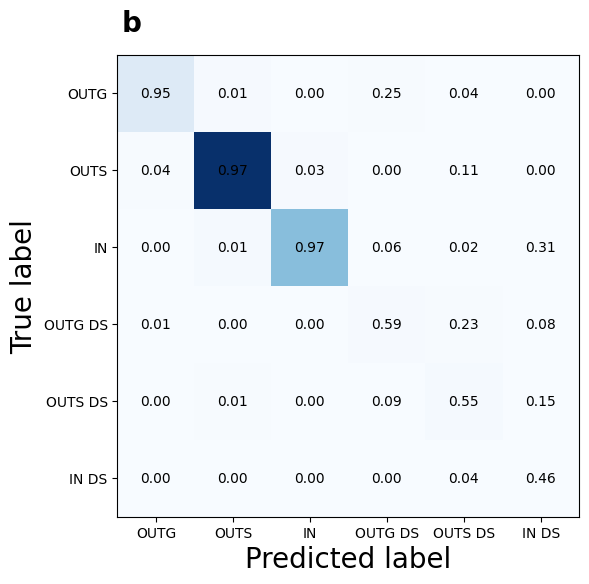}
		\caption{}
		\label{fig:lgb6:b}
	\end{subfigure}
	\begin{subfigure}[t]{0.3\textwidth}
		\centering
		\includegraphics[width=\textwidth]{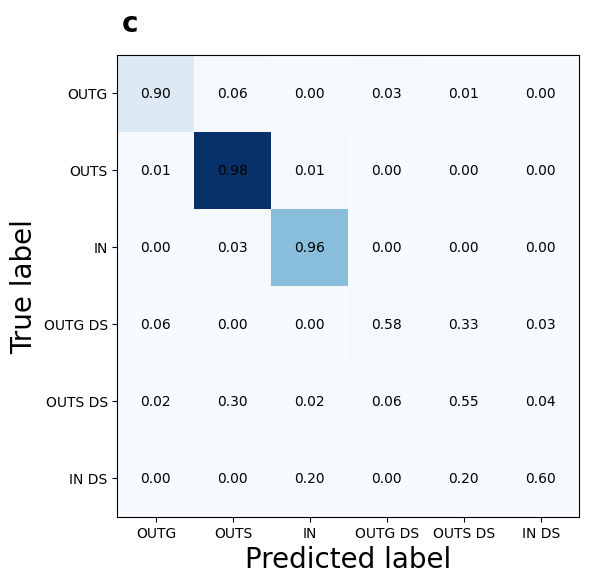}
		\caption{}
		\label{fig:lgb6:c}
	\end{subfigure}
	\begin{subfigure}[t]{0.8\textwidth}
		\centering
		\includegraphics[width=\textwidth]{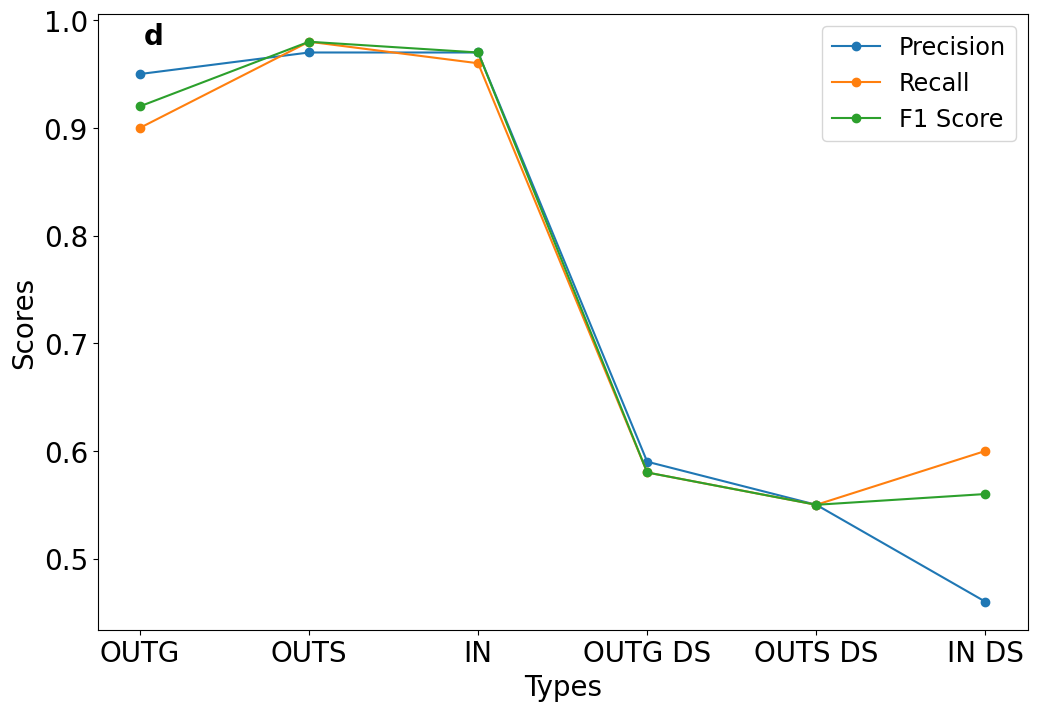}
		\caption{}
		\label{fig:lgb6:d}
	\end{subfigure}
	\caption{Panel A: The matrix displays the number of samples in the test set, with true labels represented on the y-axis and predicted labels on the x-axis. Panel B: The efficiency matrix for lightGBM, calculated by normalizing each column of the left panel to the total number of samples in that column. The values in each box correspond to the proportion of samples in the test set with labels assigned on the x-axis that belong to the classes on the y-axis, resulting in accuracy values along the diagonal. Darker colors in the three panels correspond to higher proportions or numbers of samples. Panel C: The confusion matrix for lightGBM, computed by normalizing each row of the left panel to the total number of stars in that row. The values correspond to the scores of samples in the test set with true labels displayed on the y-axis, and these samples are assigned labels on the x-axis, resulting in scores for correct classification of each class along the diagonal. In Panel D, the horizontal axis represents different types, while the vertical axis represents the scores. Blue, orange, and green correspond to precision, recall, and F1 score, respectively.
		\label{fig:lgb6}}
\end{figure*}

We utilized the lightGBM classifier to analyze the six types separately. Additionally, we defined precision, recall, and F1 score, and the principles of these metrics are presented in appendix \ref{sec:F1}. The confusion matrices, precision, recall, and F1 scores for the test set are shown in Figure \ref{fig:lgb6}. Among the six types, the precision, recall, and F1 scores for the first three types that did not experience long-term double synchronization are high, all exceeding 0.9, indicating that the model effectively distinguishes these types. However, for the last three types that experienced long-term double synchronization, despite using oversampling to balance the minority class samples, the classification results remain unsatisfactory, with all metrics below 0.6. Nonetheless, the accuracy of the classifier at this stage is high, reaching 95.78$\%$. For the three types that experienced long-term double synchronization, the poor classification results are reasonable due to the limited number of samples for each type.

\begin{figure*}
	\centering
	\begin{subfigure}[t]{0.3\textwidth}
		\centering
		\includegraphics[width=\textwidth]{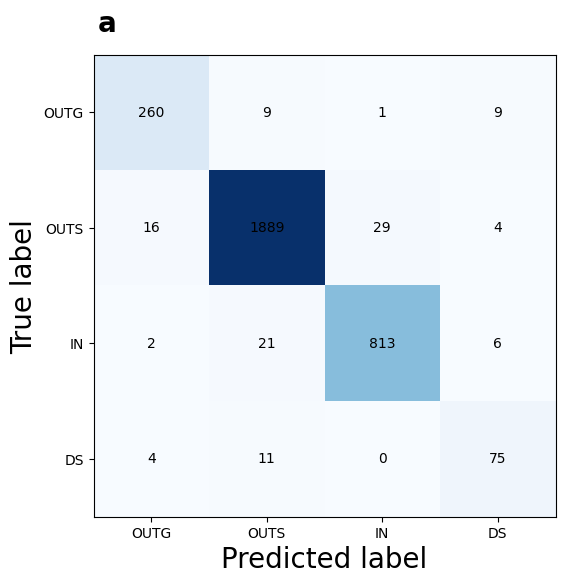}
		\caption{}
		\label{fig:lgb4:a}
	\end{subfigure}
	\begin{subfigure}[t]{0.3\textwidth}
		\centering
		\includegraphics[width=\textwidth]{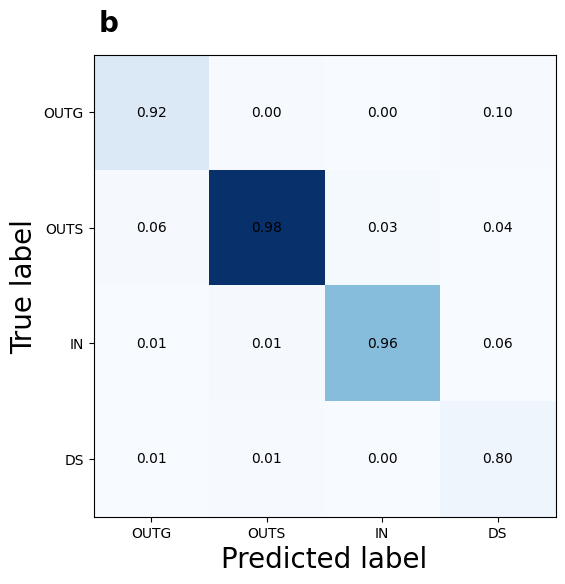}
		\caption{}
		\label{fig:lgb4:b}
	\end{subfigure}
	\begin{subfigure}[t]{0.3\textwidth}
		\centering
		\includegraphics[width=\textwidth]{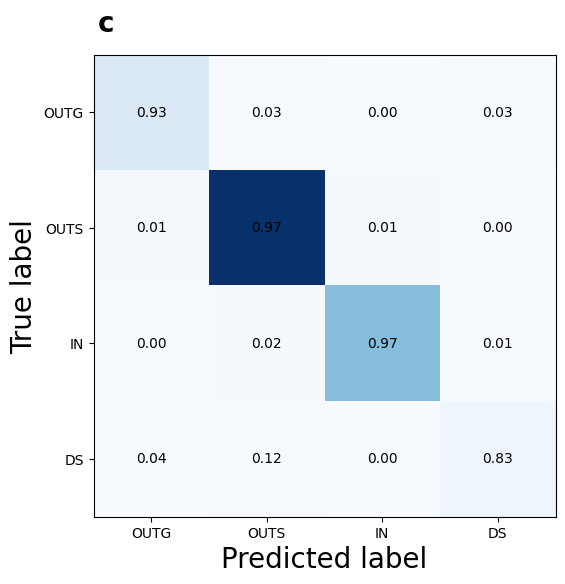}
		\caption{}
		\label{fig:lgb4:c}
	\end{subfigure}
	\begin{subfigure}[t]{0.8\textwidth}
		\centering
		\includegraphics[width=\textwidth]{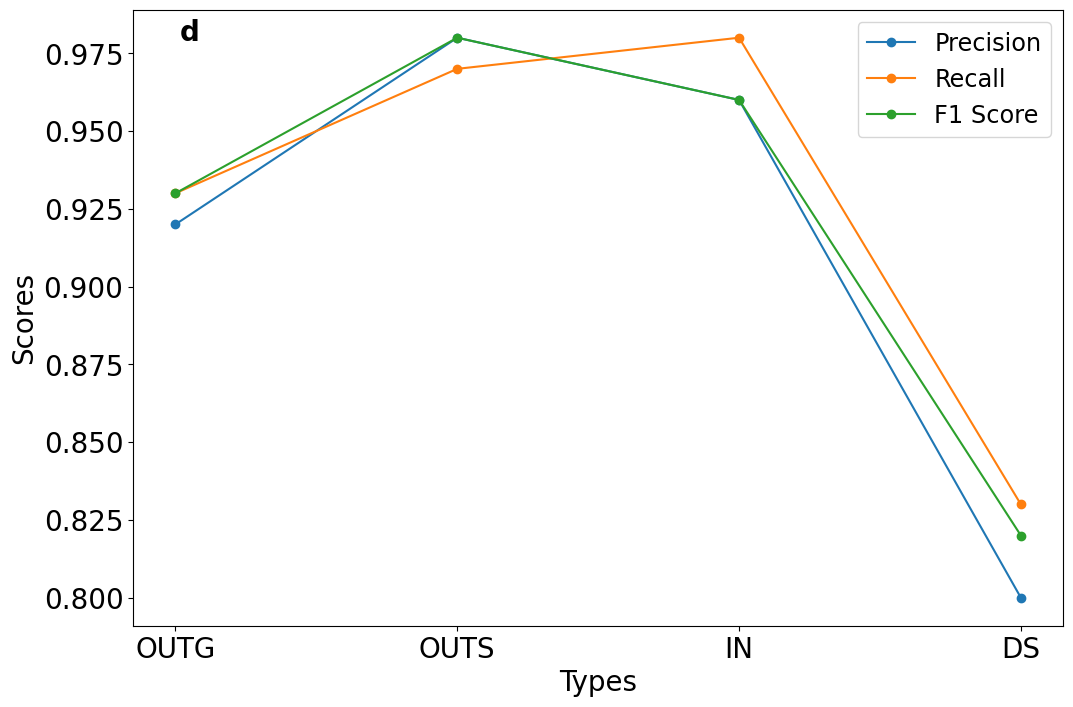}
		\caption{}
		\label{fig:lgb4:d}
	\end{subfigure}
	\caption{Similar to Figure \ref{fig:lgb6}, but with the three long-term double synchronization types combined into one type, the results of training four types using lightGBM are shown. It is worth noting that for the long-term double synchronization samples, there are evidently more samples falling on the diagonal of Panel A, Panel B and Panel C, which reflects the improved performance of lightGBM in predicting the four types.}
	\label{fig:lgb4}
\end{figure*}

To improve the model's accuracy, we merged these three types into a single class, resulting in four coarse labels. From Figure \ref{fig:lgb4}, we conducted classification on these four classes and found a significant improvement in the metrics for the long-term double synchronization types, all surpassing 0.8. The accuracy of the classifier also slightly increased to 96.44$\%$.

Finally, when given any star-planet system with the six initial input parameters, we can not only predict the system's evolutionary curve but also identify its migration state. This provides a clearer understanding of the physical significance of these systems. The importance of classification is further demonstrated by the ability to compare the evolutionary curve errors of systems with different migration states and discuss the possible reasons for these differences. This will be discussed in the next subsection.

\subsection{Error Statistics and Analysis of Rotation Period and Orbital Period for Individual Star-Planet Systems} \label{subsec:style3}
In the previous two sections, we discussed the absolute errors of all samples in the test set and the predictions of curve types. We calculated the average relative errors for each of the four evolutionary quantities in each system and obtained their median values, which are presented in Table \ref{tab:Parameter1}. We found that, except for $P_{\mathrm{rot}}$, the average relative errors of the other three evolutionary quantities are all below 1$\%$.

Figure \ref{fig:regression2} shows the $45^{\circ}$ lines of the test set for the stellar rotation period ($P_{\mathrm{rot}}$) and planetary orbital period ($P_{\mathrm{orb}}$). Although the overall errors of the samples are small, there are a few samples with significant errors. In practical applications, our goal is to obtain the evolution curve of any given system within the parameter range. Therefore, we need to analyze the errors of individual star-planet systems  predicted to evaluate when our model's predictions have relatively small errors and when they have relatively large errors. Firstly, when we employ neural networks for regression, we start by examining the relationship between inputs and periods. Specifically, we discuss the influence of six initial input parameters on the distribution of errors. Secondly, these six initial input parameters alone may not adequately assess the model's performance. We need to explore additional dimensions to evaluate the model's performance because the complexity of evolutionary curves may also impact the regression error. Therefore, we categorize the curves according to different physical laws, which in a sense, classifies the complexity of the evolutionary curves. This classification model will be used to measure the performance of our regression model. In other words, we discuss the distribution of errors based on the types of planetary migration states. For example, for the minority class samples that undergo long-term double synchronization, we can label them, considering that the regression curves of these samples have larger errors.

We evaluate the accuracy of individual systems from two aspects. First, we will discuss the impact of the six initial input parameters on the distribution of errors. Second, we will examine the distribution of errors based on the planetary survival status and types of planetary migration states.

First, we selected the curves in the test set with median relative errors of 5$\%$ to 10$\%$ and greater than 10$\%$ and labeled them as "moderate errors" and "high errors", respectively. The number of curves with moderate and high errors for $P_{\mathrm{rot}}$ and $P_{\mathrm{orb}}$ is provided in Table \ref{table3:number}. It can be found that there are a considerable number of curves with moderate and high errors for $P_{\mathrm{rot}}$, but the total does not exceed 20$\%$. In the case of $P_{\mathrm{orb}}$, the number is even smaller, accounting for less than 3$\%$, indicating that the vast majority of our predicted curves perform well.

\begin{figure*}
	\includegraphics[width=0.95\textwidth]{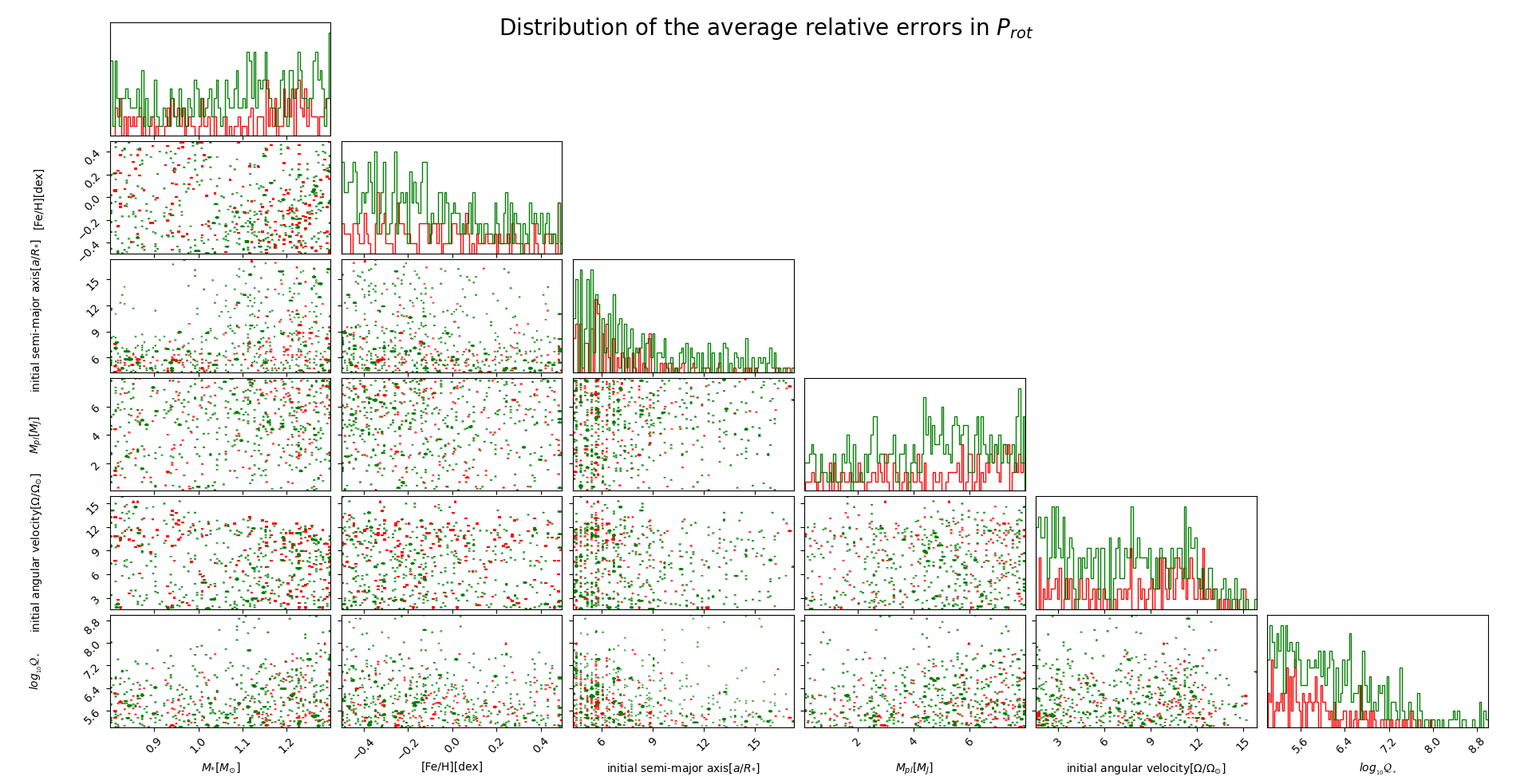}
	\caption{The figure shows the distribution of high errors and moderate errors of stellar rotation period $P_{\mathrm{rot}}$ with respect to six initial parameters. High errors are represented in red, and moderate errors are represented in green.   \label{fig:DPR}}
\end{figure*}

\begin{figure*}
	\includegraphics[width=0.95\textwidth]{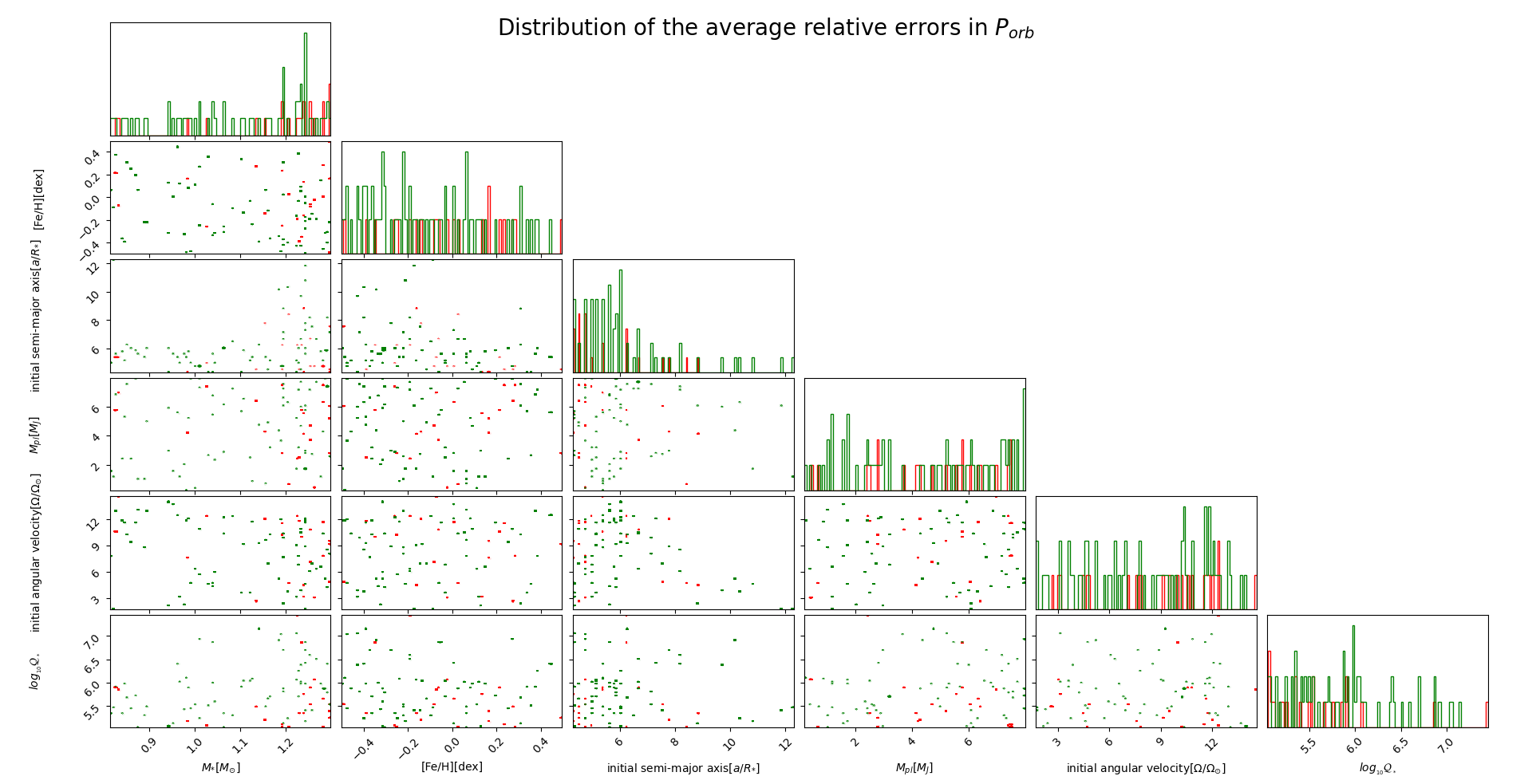}
	\caption{The figure shows the distribution of high errors and moderate errors of planetary orbital period $P_{\mathrm{orb}}$ with respect to six initial parameters. High errors are represented in red, and moderate errors are represented in green.   \label{fig:DPO}}
\end{figure*}

From Figures \ref{fig:DPR} and \ref{fig:DPO}, it can be seen that curves with errors greater than 5$\%$ and 10$\%$ are mainly concentrated in systems with characteristics such as high mass, low metallicity, low initial semi-major axis, and low tidal quality factor. Our predictions are more sensitive to the initial semi-major axis and tidal quality factor. This is because high mass, low metallicity, close-in orbits, and low tidal quality factor imply faster orbital migration rates for the planet. Rapid orbital evolution is accompanied by faster angular momentum transfer between the orbit and the star, leading to faster changes in stellar rotation. This results in steeper curves and larger prediction errors. Furthermore, the initial semi-major axis and tidal quality factor are the most influential factors on the planet's orbital migration, making them the most sensitive parameters.

\begin{figure*}
	\includegraphics[width=0.95\textwidth]{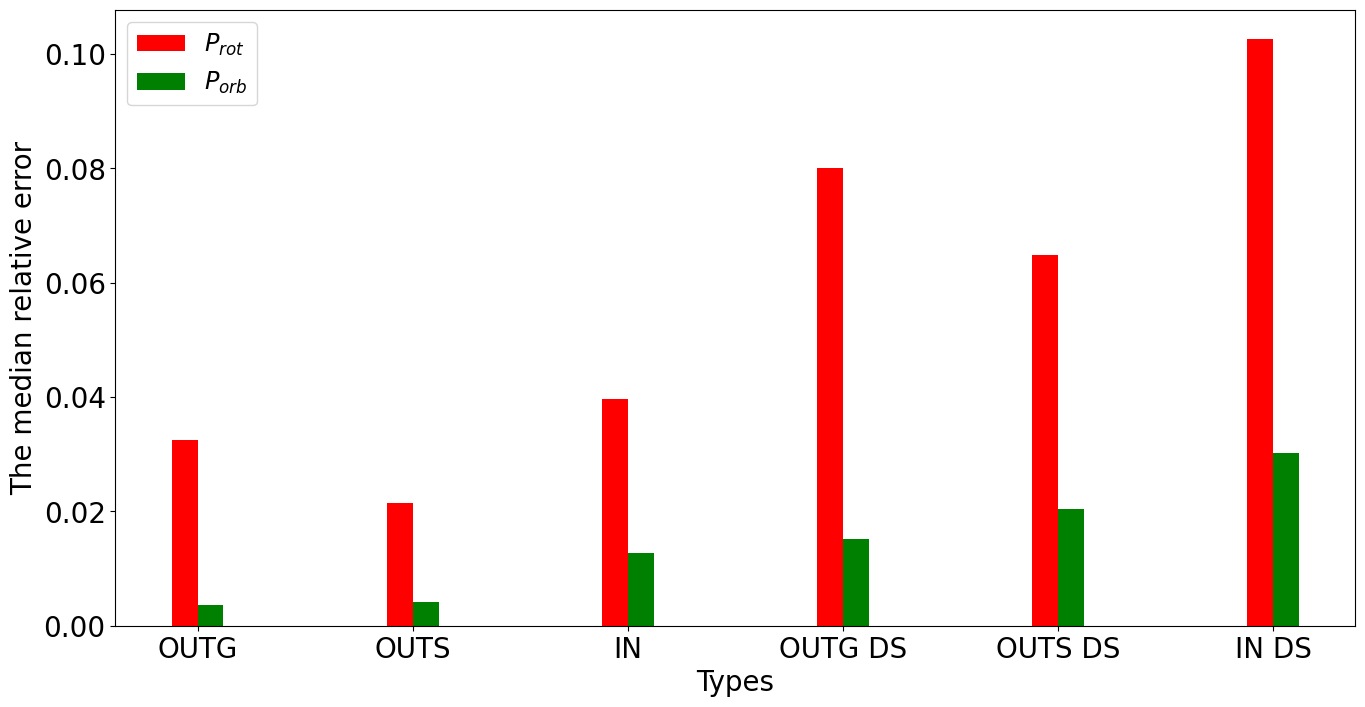}
	\caption{The figure presents the median of the average relative errors for each curve under different types of $P_{\mathrm{rot}}$ and $P_{\mathrm{orb}}$. It can be observed that the errors are significantly higher for systems experiencing long-term double synchronization.   \label{fig:CE}}
\end{figure*}

Subsequently, from Figure \ref{fig:CE}, it can be found that both $P_{\mathrm{rot}}$ and $P_{\mathrm{orb}}$ have median relative errors for systems experiencing long-term double synchronization. This may be attributed to two factors. Firstly, there are very few samples for these systems, accounting for less than 3$\%$ of the total dataset, which does not provide sufficient data for the model to be adequately trained. Secondly, due to the specific nature of these curves, as evident from Figure \ref{fig:type} (d), (e), and (f), the most prominent feature of this type is the slow decay of both the planetary orbital period and the stellar rotation period when they become equal. The complexity of these evolution curves may potentially affect the precision of the model training.

\begin{figure*}
	\centering
	\begin{subfigure}[t]{0.15\textwidth}
		\centering
		\includegraphics[width=\textwidth]{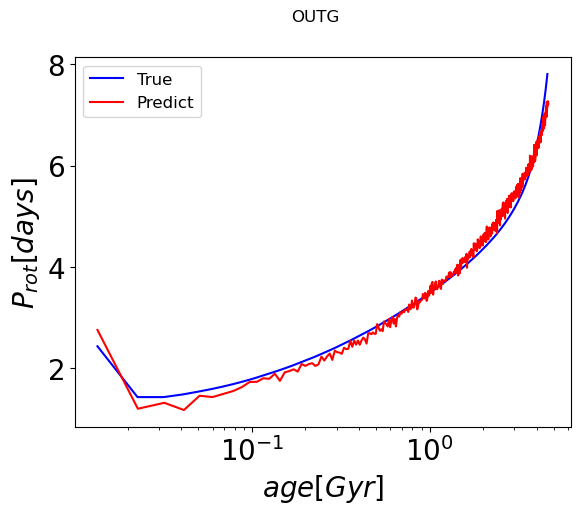}
		\caption{}
		\label{fig:curve:a}
	\end{subfigure}
	\begin{subfigure}[t]{0.15\textwidth}
		\centering
		\includegraphics[width=\textwidth]{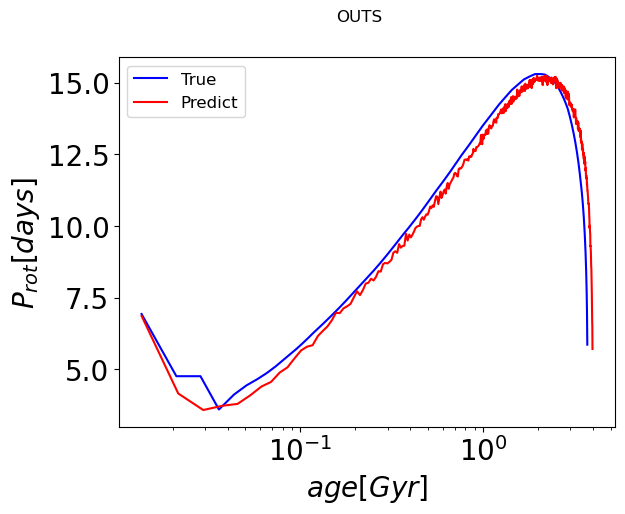}
		\caption{}
		\label{fig:curve:b}
	\end{subfigure}
	\begin{subfigure}[t]{0.15\textwidth}
		\centering
		\includegraphics[width=\textwidth]{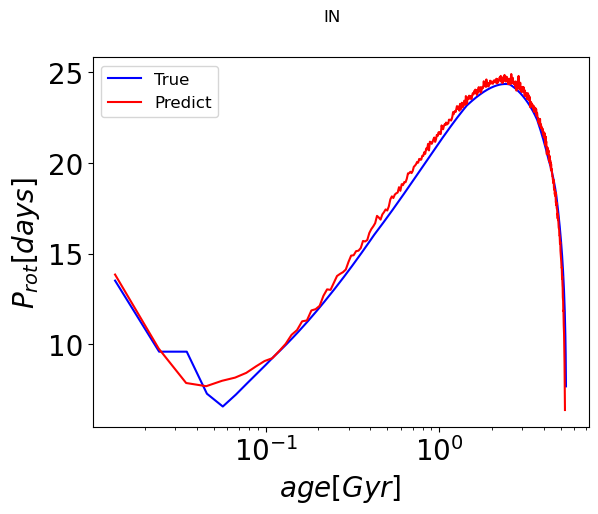}
		\caption{}
		\label{fig:curve:c}
	\end{subfigure}
	\begin{subfigure}[t]{0.15\textwidth}
		\centering
		\includegraphics[width=\textwidth]{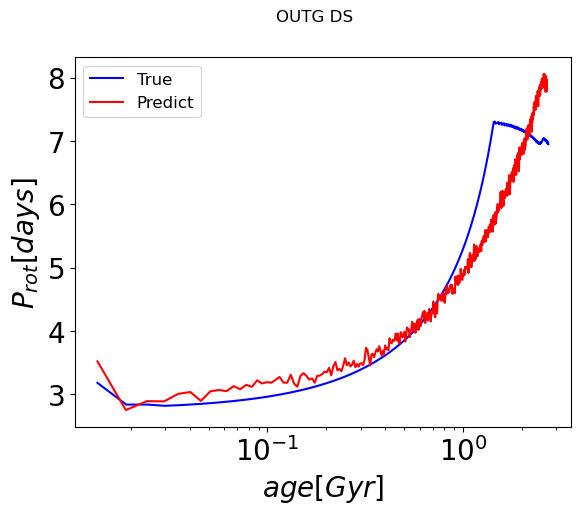}
		\caption{}
		\label{fig:curve:d}
	\end{subfigure}
	\begin{subfigure}[t]{0.15\textwidth}
		\centering
		\includegraphics[width=\textwidth]{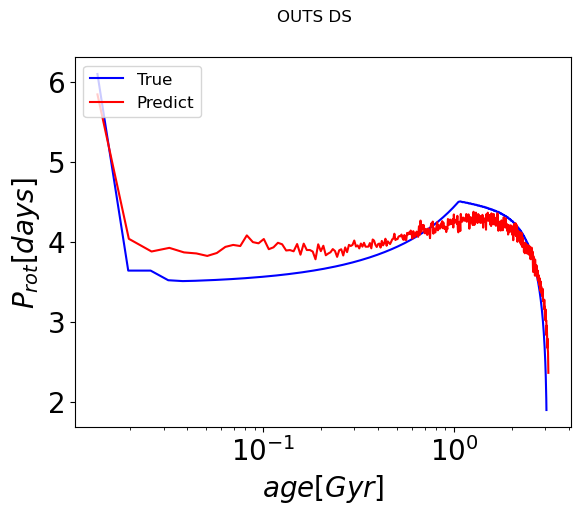}
		\caption{}
		\label{fig:curve:e}
	\end{subfigure}
	\begin{subfigure}[t]{0.15\textwidth}
		\centering
		\includegraphics[width=\textwidth]{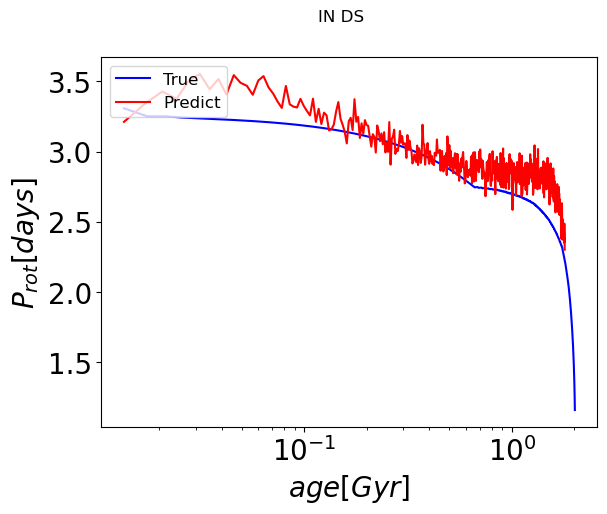}
		\caption{}
		\label{fig:curve:f}
	\end{subfigure}
	\begin{subfigure}[t]{0.15\textwidth}
		\centering
		\includegraphics[width=\textwidth]{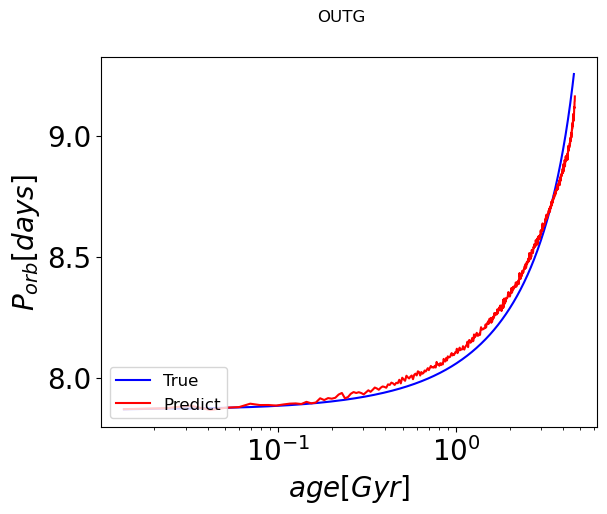}
		\caption{}
		\label{fig:curve:g}
	\end{subfigure}
	\begin{subfigure}[t]{0.15\textwidth}
		\centering
		\includegraphics[width=\textwidth]{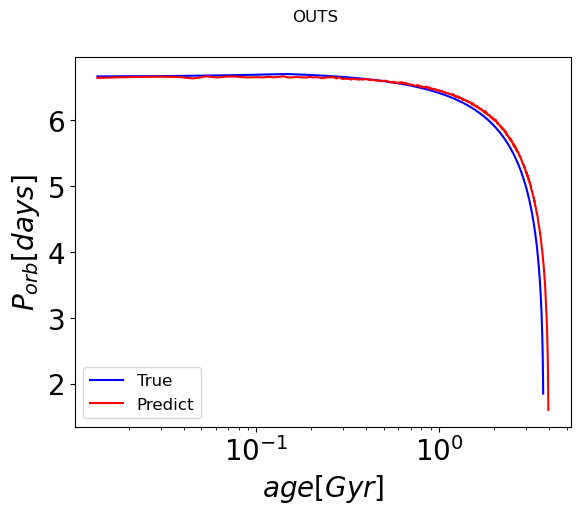}
		\caption{}
		\label{fig:curve:h}
	\end{subfigure}
	\begin{subfigure}[t]{0.15\textwidth}
		\centering
		\includegraphics[width=\textwidth]{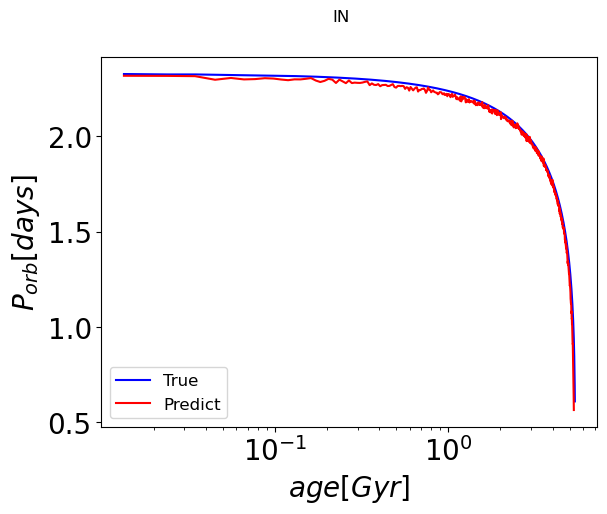}
		\caption{}
		\label{fig:curve:i}
	\end{subfigure}
	\begin{subfigure}[t]{0.15\textwidth}
		\centering
		\includegraphics[width=\textwidth]{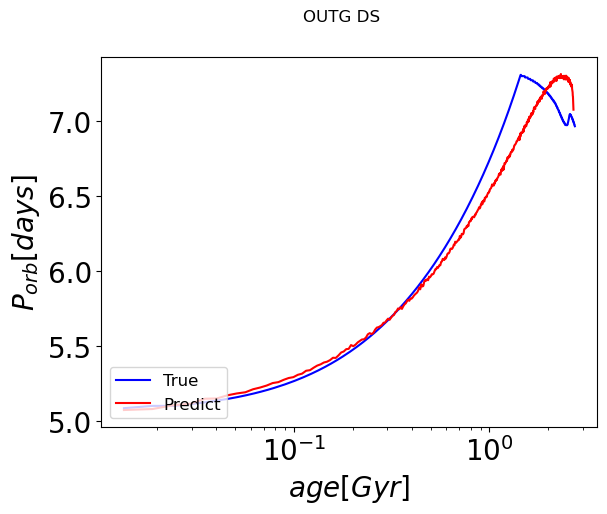}
		\caption{}
		\label{fig:curve:j}
	\end{subfigure}
	\begin{subfigure}[t]{0.15\textwidth}
		\centering
		\includegraphics[width=\textwidth]{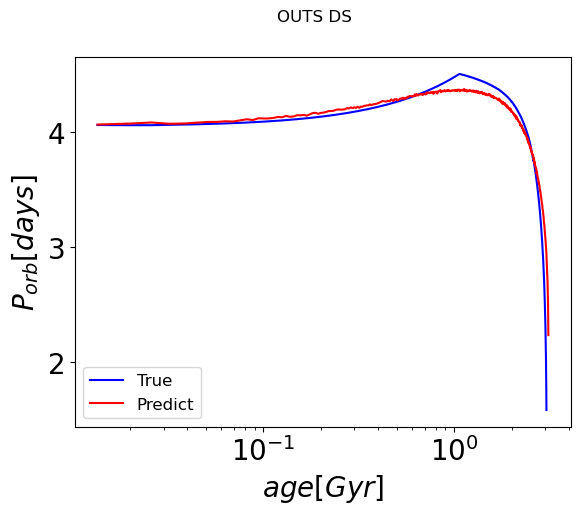}
		\caption{}
		\label{fig:curve:k}
	\end{subfigure}
	\begin{subfigure}[t]{0.15\textwidth}
		\centering
		\includegraphics[width=\textwidth]{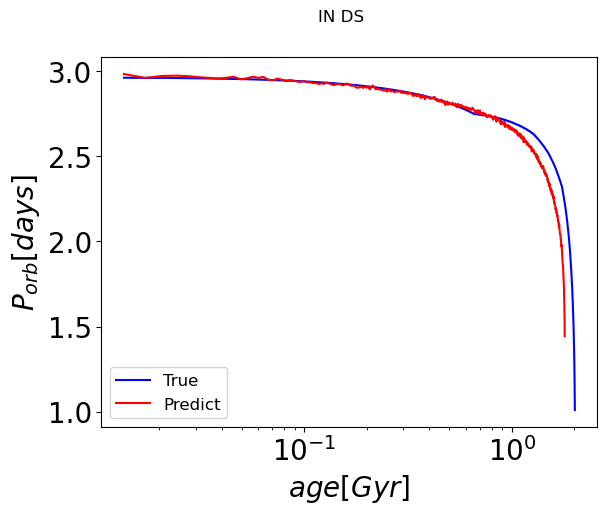}
		\caption{}
		\label{fig:curve:l}
	\end{subfigure}
	\caption{
		The figure presents complete examples of the evolutionary curves for the rotational period ($P_{\mathrm{rot}}$) and orbital period ($P_{\mathrm{orb}}$) corresponding to the six migration states. The blue solid lines represent the original evolutionary curves, while the red solid lines depict the predicted evolutionary curves. The upper part of each subplot shows the stellar rotational period, while the lower part displays the planetary orbital period. The subplots are arranged from left to right as follows: a. OUTG, b. OUTS, c. IN, d. OUTG DS, e. OUTS DS, and f. IN DS. The specific parameters for each migration state are presented in Table \ref{tab:Parameter1}.
		\label{fig:curve}}
\end{figure*}

Finally, in Figure \ref{fig:curve}, we present examples of the rotational period ($P_{\mathrm{rot}}$) and orbital period ($P_{\mathrm{orb}}$) from the complete evolutionary curves for each of the six types. Overall, our predicted evolutionary curves do not deviate significantly from the actual curves. However, as analyzed earlier, the three types undergoing long-term double synchronization exhibit larger deviations, particularly the "IN DS" type, which has the smallest sample size and the largest deviation in stellar rotational period ($P_{\mathrm{rot}}$). Consequently, for systems undergoing long-term double synchronization, our classifier can correctly identify them with an accuracy of 80$\%$. However, caution is warranted when using the predicted evolutionary curves for such samples due to their lower accuracy.

\subsection{The Speed of Generating Evolutionary Curves and Features for Star-Planet Systems} \label{subsec:speed}
Our model significantly accelerates the generation speed of theoretical curves. For instance, the time required to generate a single evolutionary curve for an evolutionary parameter is only 0.015 seconds, and for four evolutionary parameters, it is only 0.06 seconds. This is merely five ten-thousandths of the time required for numerical computation relying solely on theoretical models. For the 20\% test set (3149 models), predicting the "maximum age" and type of all models only takes 0.02 seconds, which is negligible. Thus, our model significantly speeds up the generation of evolutionary curves and features, saving computational resources and time.

\begin{table}
	\centering
	\caption{The median relative errors of the evolutionary quantities.}
	\label{tab:Parameter1}
	\begin{tabular}{cccc}
		\hline
		 $log_{T_{eff}}$ & $R_*$ & $P_{\mathrm{rot}}$ & $P_{\mathrm{orb}}$ \\
		\hline
	      0.151$\%$ & 0.431$\%$ & 2.609$\%$ & 0.571$\%$ \\
		\hline
	\end{tabular}
\end{table}

\begin{table}

	\caption{The number of high errors and moderate errors}
	\centering
	\begin{tabular}{|c|c|c|c|c|}
		\hline
		& \multicolumn{2}{c|}{$P_{\mathrm{rot}}$} & \multicolumn{2}{c|}{$P_{\mathrm{orb}}$} \\
		\cline{2-5}
		& moderate errors & high errors & moderate errors & high errors \\
		\hline
		Number & 413 & 166 & 67 & 22 \\
		\hline
	\end{tabular}
	\label{table3:number}
\end{table}

\begin{table}

	\caption{The parameters adopted in Figure \ref{fig:curve} \label{tab:Parameter2}}
	\centering
	\begin{tabular}{|l|c|c|c|c|c|c|}
		\hline
		migration states & $M_{*}$ ($M_{\odot}$) & $\mathrm{\mathrm{\mathrm{[Fe/H]}}}$ (dex) & $P_\mathrm{orb, ini}$ (days) & $M_\mathrm{pl}$ ($M_\mathrm{J}$) &
		$P_\mathrm{rot, ini}$ (days) & $logQ_{*}$ \\
		\hline
		OUTG & 1.062 & -0.425 & 7.903 & 3.099 & 2.597 & 5.185 \\
		\hline
		OUTS & 0.925 & -0.346 & 7.990 & 6.169 & 1.810 & 6.061 \\
		\hline
		IN & 1.064 & 0.271 & 1.756 & 1.992 & 4.005 & 7.771 \\
		\hline
		OUTG DS & 1.229 & -0.316 & 5.108 & 4.726 & 3.381 & 5.009 \\
		\hline
		OUTS DS & 1.204 & -0.172 & 4.079 & 3.864 & 6.721 & 6.038 \\
		\hline
		IN DS & 1.294 & -0.418 & 2.976 & 7.287 & 3.328 & 6.923 \\
		\hline
	\end{tabular}
\end{table}

\section{CONLUSION AND DISCUSSION} \label{sec:CAD}
We used the MESA to model the stellar-planet interaction and generated 7500 single-star samples and 15745 star-planet system samples within a certain parameter space. We employed the neural network model MLP for regression of our evolution parameters. To achieve higher precision, we conducted regression using the single-star samples for the effective temperature $logT_\mathrm{eff}$, stellar radius $R_*$, and maximum main-sequence age of stars. For the star-planet system samples, we conducted regression for the stellar rotation period, planetary orbital period, and maximum age of the system.
The median relative errors for the effective temperature $logT_\mathrm{eff}$, stellar radius $R_*$, stellar rotation period $P_{\mathrm{rot}}$, and planetary orbital period $P_{\mathrm{orb}}$ were 0.151$\%$, 0.431$\%$, 2.609$\%$, and 0.571$\%$, respectively. Our results indicate that for any given set of six initial input parameters, we can obtain complete time-evolving plots of the four evolution parameters using MLP predictions. Additionally, the precision obtained from the transit method, which directly measures the time intervals between each transit, is quite high. Except for the planetary orbital period, the errors in the other evolution parameters are significantly smaller than the observational errors. This suggests that using machine learning methods to rapidly generate evolution curves not only greatly reduces the computation time compared to using physical models but also ensures high accuracy.

We also classified the samples into 6 types based on the planetary migration state. Using lightGBM, we performed oversampling on the three minority class samples that experienced long-term double synchronization. Additionally, we applied stratified 10-fold cross-validation and voting on the training set for all types to reduce the model performance fluctuations caused by the randomness of dataset partitioning and improve the model's stability. However, we found that the prediction results for the three minority class samples experiencing long-term double synchronization remained poor. This could be attributed to the limited number of samples and the uniqueness of the curves. Yet, when we merged the three classes of long-term double synchronization into one type, we found that the four types were clearly distinguishable. This indicates that our model is capable of distinguishing these types effectively. Our ultimate goal is to use MLP to generate complete evolutionary curves. To assess the accuracy of each generated evolutionary curve, we focused on two parameters, $P_{\mathrm{rot}}$ and $P_{\mathrm{orb}}$, which showed higher errors. We found that the main sources of errors were systems with larger stellar and planetary masses, lower stellar metallicity, and smaller initial orbital semi-major axis and tidal quality factor. Additionally, for the planetary migration state, the main source of errors originated from the samples experiencing long-term double synchronization.

In conclusion, our work provides a tool that allows us to regress the physical evolutionary quantities of interest using only six initial parameters and generate complete evolutionary profiles. We do not need to spend a significant amount of time and resources calculating theoretical models but instead use machine learning to train a large database within the parameter space for any star-planet systems model. Additionally, we propose the feasibility of using machine learning methods to explore theoretical data from a large number of stellar-planet interaction models. We can classify different star-planet systems based on physical processes and train machine learning models to determine the weights of different input parameters in each type. This provides a classifier that makes these physical processes clearly distinguishable. We also acknowledge the limitations of our work. Firstly, our model does not consider the influence of planetary magnetic fields and mass loss due to photoevaporation on planetary orbital migration \citep{2010ApJ...712.1107G}, which has been explored in previous theoretical works. Furthermore, some initial quantities are uncertain in observations, such as the tidal quality factor and the initial planetary orbital distance and stellar initial rotation frequency. We hope to combine observational data to explore the range and relationships of these initial physical quantities in real star-planet systems in future work. The large database provided by our current work will pave the way for future implementations. We also consider building more comprehensive interaction models and using machine learning to explore larger parameter spaces and more suitable physical parameters.
		
		software:MESA R11554 \citep{ 2011ApJS..192....3P,2013ApJS..208....4P,2015ApJS..220...15P,2018ApJS..234...34P,2019ApJS..243...10P}
		
		\section*{Acknowledgements}
		I am grateful to the reviewers for their insightful feedback and valuable contributions to this manuscript. This work is supported by the Strategic Priority Research Program of Chinese Academy of Sciences, Grant No. XDB 41000000 and National Natural Science Foundation of China (Nos. 12288102). J. Guo 	  		acknowledges the support by the National Natural Science Foundation of China (No. 11973082). D. Yan acknowledges the support by the National Natural Science Foundation of China (No. 42305136). This work is also supported by the National Key R\&D Program of China (grant No. 2021YFA1600400/2021YFA1600402), 		Natural Science Foundation of Yunnan Province (No. 202201AT070158) and the International Centre of Supernovae, Yunnan Key Laboratory (No.202302AN360001). The authors gratefully acknowledge the “PHOENIX Supercomputing Platform” jointly operated by the Binary Population Synthesis Group and the 					Stellar Astrophysics Group at Yunnan Observatories, Chinese Academy of Sciences. KaiFan Ji and Hui Liu were partly supported by Yunnan KeyLaboratory of Solar Physics and Space Science under the number 202205AG070009.
		\section{Data availability}
		The data underlying this article will be shared on reasonable request to the corresponding author.
		
		\appendix
		\section{Evaluation Metrics for Machine Learning Classification Models}  \label{sec:F1}

		Precision, Recall, and F1 Score are commonly used metrics to assess the performance of classification models. To evaluate the classification performance of our model, we introduce these metrics and present their expressions and meanings as follows:
		
		Precision: Precision is the proportion of samples predicted as positive that are actually positive, measuring the accuracy of the model's positive predictions.
		
		$\text{Precision} = \frac{{\text{TP}}}{{\text{TP + FP}}}$
		
		Recall: Recall is the proportion of actual positive samples that are correctly predicted as positive, measuring the model's ability to identify positive samples.
		
		$\text{Recall} = \frac{{\text{TP}}}{{\text{TP + FN}}}$
		
		F1 Score: F1 Score is the weighted harmonic mean of precision and recall, used to provide a comprehensive evaluation of the model's accuracy and recall capabilities.
		
		$\text{F1 Score} = \frac{{2 \times (\text{Precision} \times \text{Recall})}}{{\text{Precision + Recall}}}$
		
		Here, TP represents true positives, TN represents true negatives, FP represents false positives, and FN represents false negatives.

		
		
		\bibliographystyle{mnras}
		\bibliography{example.bib} 

		
		
		\onecolumn

		\bsp	
		\label{lastpage}
\end{document}